\documentclass{aa}
\usepackage{psfig}
\usepackage{txfonts}
\usepackage{natbib}
\usepackage{amssymb}
\bibpunct{(}{)}{;}{a}{}{,}

\def\pcsk{ph~cm$^{-2}$s$^{-1}$keV$^{-1}$}
\def\ecs{erg~cm$^{-2}$s$^{-1}$}
\def\es{erg\,s$^{-1}$}

\def\chir{$\chi^2_r$ }

\def\igr{INTEGRAL}
\def\xmm{XMM-{\it Newton}}

\def\pca{RXTE-PCA}

\def\sgr1806{SGR~1806-20}
\def\sg1900{SGR~1900+14}

\def\0142{4U~0142+61}
\def\axp1048{1E~1048.1-5937}
\def\e1547{1E~1547.0-5408}

\def\rxs1708{1RXS~J1708-40}
\def\xte1810{XTE~J1810-197}
\def\1841{1E~1841-045}
\def\ax1845{AX~J1845.0-0258}
\def\2259{1E~2259+586}

\begin{document}
\bibliographystyle{aa}

\title{Detailed high-energy characteristics of AXP 1RXS~J170849-400910}
\subtitle{Probing the magnetosphere using \igr, RXTE and \xmm}

\author{P.R. den Hartog\inst{1}
  \and L. Kuiper\inst{1}
  \and W. Hermsen\inst{1,2}}

\offprints{P.R. den Hartog}
\mail{Hartog@sron.nl}

\institute{SRON, Netherlands Institute for Space Research,
  Sorbonnelaan 2, 3584 CA Utrecht, The Netherlands
  \and Sterrenkundig Instituut Anton Pannekoek, University of Amsterdam,
  Kruislaan 403, 1098 SJ Amsterdam, The Netherlands}

\date{Received 13 March 2008 / Accepted ...}

\abstract{1RXS~J170849-400910 is one of four Anomalous X-ray Pulsars
  which emit persistent luminous radiation in soft X-rays ($<$10~keV)
  as well as in hard X-rays ($>$10~keV). In this paper we present
  detailed spectral and temporal characteristics over the whole X-ray
  band. For this purpose data have been used from \igr, RXTE and \xmm.
  The hard X-ray ($>$10~keV) time-averaged total spectrum, accumulated
  over four years with the imager IBIS-ISGRI onboard \igr\ adding up
  to 5.2 Ms net exposure, can be described by a power law with a
  photon index $\Gamma = 1.13 \pm 0.06$ and extends to
  $\sim$175~keV. The 20--175~keV flux is $(7.76 \pm 0.34) \times
  10^{-11}$\,\ecs which exceeds the 2--10~keV (unabsorbed) flux by a
  factor of $\sim$2.3. No evidence for a spectral break is found below
  300~keV. Also, no significant long-term time variability has been
  detected above 20~keV on time scales of 1 and 0.5 year. Pulsed
  emission is measured with \igr\ up to 270~keV, i.e. to much higher
  energies than the total emission, with a detection significance of
  12.3$\sigma$ (20--270~keV). The pulse profiles from 0.5~keV up to
  270~keV show drastic morphology changes below $\sim$20~keV. Three
  different pulse components can be recognized in these pulse
  profiles: 1) a hard pulse peaking around phase 0.8 which contributes
  to the pulse profiles above $\sim$4~keV, 2) a softer pulse which
  peaks around phase 0.4 not contributing in the hard X-ray domain and
  3) a very soft pulse component below 2~keV.  A combined
  time-averaged pulsed spectrum (2.8--270~keV) from \igr, \pca\ and
  HEXTE (collected over nine years) can be described with a soft and a
  hard power-law component: $\Gamma_{{\rm s}}= 2.79 \pm 0.07$ and
  $\Gamma_{{\rm h}}= 0.86 \pm 0.16$. In the pulsed spectrum extracted
  from a 25.5~ks net exposure \xmm\ observation we find a
  discontinuity between 2 keV and 3 keV.  Above these energies the
  spectrum is consistent with the spectrum taken with \pca. The pulse
  profiles and the total-pulsed spectrum prove to be stable over the
  whole nine-years time span over which the data have been taken. Also
  detailed phase-resolved spectroscopy of the pulsed emission confirms
  the long-term stability as the spectra taken at different epochs
  connect smoothly. The phase-resolved spectra reveal complex spectral
  shapes which do not follow the shape of the total-pulsed
  spectrum. The spectral shape gradually changes with phase from a
  soft single power law to a complex multi-component shape and then to
  a hard single power law. The spectrum switches from a very hard
  ($\Gamma = 0.99 \pm 0.05$) to a very soft ($\Gamma = 3.58 \pm 0.34$)
  single power-law shape within a 0.1-wide phase interval. The
  discontinuity measured between 2~keV and 3~keV with \xmm\ is a
  result of a curved component. This component which is most apparent
  within phase interval 0.7--0.9 significantly contributes in the
  energy range between 4~keV and 20~keV. It has a very steep spectrum
  below 5~keV with a photon index $\Gamma \sim -1.5$. From the
  phase-resolved spectra we identify three independent components with
  different spectral shapes which together can accurately describe all
  phase-resolved spectra (2.8--270~keV). The three shapes are a soft
  power law ($\Gamma = 3.54$), a hard power law ($\Gamma = 0.99$) and
  a curved shape (described with two logparabolic functions). The
  phase distributions of the normalizations of these spectral
  components form three decoupled pulse profiles. The soft component
  peaks around phase 0.4 while the other two components peak around
  phase 0.8. The width of the curved component ($\sim$0.25 in phase)
  is about half the width of the hard component. After \0142, \rxs1708
  is the second anomalous X-ray pulsar for which such detailed
  phase-resolved spectroscopy has been performed. These results give
  important constraints showing that three dimensional modeling
  covering both the geometry and different production processes is
  required to explain our findings.

\keywords{ -- Stars: neutron -- X-rays: individuals:
  \object{1RXS~J170849-400910},
  \object{\0142} -- Gamma rays:
  observations

}}

\maketitle
\section{Introduction}
\label{sec:rxsintro}

In the past five years, several new developments have changed our view
on Anomalous X-ray Pulsars (AXPs) drastically \citep[see for a recent
  review][]{Kaspi07_london}. AXPs are young isolated rotating neutron
stars.  Their spin periods and spin derivatives indicate that the
surface magnetic fields of AXPs range from $\sim$$6 \times 10^{13}$\,G
to $\sim$$7 \times 10^{14}$\,G (assuming a dipole field; $B = 3.2
\times 10^{19} \sqrt{P \dot{P}}$\,G).  The X-ray luminosity of
persistent AXPs can be orders of magnitude higher than explained by
spin-down power. Nowadays, it is generally accepted that AXPs are
powered by magnetic dissipation and that AXPs belong to the magnetar
class \citep{DT92, TD93, TD95,TD96, TLK02}. To the magnetar class also
belong Soft Gamma-ray Repeaters (SGRs) \citep{Woods06_review,
  Mereghetti07_london}. So far, there are in total 13 confirmed
magnetars known.

Traditionally, AXPs are soft X-ray sources ($\sim$0.5--10\,keV) with
thermal-like spectra ($kT \sim 0.4-0.7$\, keV) plus a soft
power-law-like component with a photon index $\Gamma \sim 3 - 4$. This
traditional view changed since the detections of extremely bright and
persistent hard X-ray emission from AXPs by
\igr\ \citep[$>$20~keV][]{Molkov04_sagarm,Revnivtsev04_gc,
  denHartog04_atel0142}.  The hard X-ray spectra of these AXPs can be
described with power-law models with photon indices $\Gamma \sim
0.9-1.4$ and are observed up to 230~keV
\citep{Kuiper04_1841,denHartog06_casa,Kuiper06_axps,denHartog08_0142}.
Because no counterparts have been found in the MeV gamma-ray domain by
CGRO-COMPTEL (0.75--30~MeV), these spectra are expected to show
breaks. For AXP \0142 \citet{denHartog08_0142} measured the first
spectral break in an AXP hard X-ray spectrum. The maximum luminosity
of \0142 is estimated to be near 280~keV.

Moreover, \citet{Kuiper04_1841, Kuiper06_axps} found that a
significant part of the hard X-ray emission is pulsed with spectral
shapes harder than the total-emission spectra. For \0142 pulsed
emission was detected up to 160 keV by \citet{denHartog08_0142}.
These authors performed for \0142 a phase-resolved spectral analysis
of the pulsed emission (i.e. without the dominating non-pulsating DC
emission) in a broad band from 0.5\,keV to 160\,keV exploiting data
from four space missions. In that study \citet{denHartog08_0142}
showed that in the pulse profiles at least three different pulse
components can be recognized which appeared to have different
spectra. Therefore, these pulse components are genuinely different
from each other. These findings provide new constraints for modelling
the underlying physical processes responsible for these spectra and
the geometry of the production sites in the AXP
magnetospheres. Theoretical models currently in development
\citep[e.g.][]{BT07, Heyl05_qed2, Baring07_london} do so far not
explain these detailed results.

In this paper we present new high-energy results from AXP
1RXS~J170849-400910 (hereafter \rxs1708). \rxs1708 has a spin period
of $P = 11.0$ s and a period derivative of $\dot{P} = 1.9 \times
10^{-11}$ ss$^{-1}$ \citep{Gavriil02_rxtemonitoring}. The
characteristic age ($\tau = P/2\dot{P}$) dipolar surface
magnetic-field strength of \rxs1708 inferred from these values are 9.0
kyr and $4.7 \times 10^{14}$ G, respectively. In the 2--10~keV band
its luminosity $L_{\rm X}$ is $1 \times 10^{35}$ \es, similar to all
other AXPs assuming a distance of 3.8 kpc
\citep{Durant06_distances}. Soon after its discovery
\citep{Sugizaki97_1708} \rxs1708\ has been included in the RXTE
monitoring program of AXPs \citep{Gavriil02_rxtemonitoring}, providing
a long base-line of observations with a wealth of new information like
timing parameters, pulsed-flux history and glitches \citep[see
  e.g.][]{Kaspi99_axps, Dib08_glitches}. For this source an
intensity--hardness correlation in the soft X-ray band ($<$10~keV) is
claimed \citep{Rea07_1708, Campana07_1708}. An indication that this
correlation also holds in the hard X-ray domain
\citep[$>$10~keV][]{Gotz07_1708} is not confirmed in this work.

\rxs1708 was first detected in the hard X-ray regime ($>$20~keV) by
\citet{Revnivtsev04_gc} in a deep \igr\ Galactic centre
map. \citet{Kuiper06_axps} discovered pulsed hard X-ray emission from
this source using \pca\ and HEXTE data. These authors also analysed
early-mission \igr\ data adding up to almost 1~Ms effective on-source
exposure and found that the hard X-ray spectrum could be
described by a power-law model with a photon index of 1.44 $\pm$
0.45. Timing analysis revealed significant pulsed emission in
the 20--300~keV band.

This work is a continuation of the approach followed in
\citet{denHartog08_0142} for AXP~\0142 applying the same methods now
for \rxs1708. We address again the total high-energy window, using
data collected over four years of \igr\ observations adding up to
$\sim$5~Ms of on-source exposure, nine years of RXTE monitoring adding
up to an exposure of $\sim$600~ks and a $\sim$25~ks
\xmm\ observation. A significantly improved time-averaged \igr\ total
spectrum is presented, as well as those for shorter ($\sim$1 year and
$\sim$0.5 year) data sets to look for possible long-term time
variability. Most importantly, we present the (time-averaged)
total-pulsed spectra as well as (time-averaged) phase-resolved pulsed
spectra from \igr, RXTE and \xmm.

\section{Observations and analysis}
\label{sec:rxsobs}

\subsection{INTEGRAL}
\label{sec:igrl}

\begin{table*}
\centering
\renewcommand{\tabcolsep}{1.7mm}
\caption[]{Summary of the INTEGRAL observations of \rxs1708. The
  following information is given: the revolution intervals, the
  observations time spans both in MJD as in calendar dates, the number
  of ScWs, exposure times ($t_{exp}$), and the effective on-source 
  exposure ($t_{eff}$).}

\begin{tabular}{l r@{ -- }l l@{ }r l@{ -- } l@{ }l l r r@{.}l  r@{.}l}

\vspace{-3mm}\\

\hline
\hline
\vspace{-3mm}\\
Rev. &
\multicolumn{8}{c}{Time span} &
ScWs &
\multicolumn{2}{r}{$t_{\mathrm{exp}}\,(\rm{ks})$} &
\multicolumn{2}{r}{$t_{\mathrm{eff}}\,(\rm{ks})$}\\
\hline
\vspace{-3mm}\\
037 -- 485 & 52668 & 54013 & Jan.&29, &2003 & Oct. &5, &2006 &
 5161 & 11959&1 & 5191&0\\
\hline
\vspace{-3mm}\\
037 -- 063 & 52668 & 52751 & Jan.&29, &2003 & Apr. &22, &2003 &
  706 & 1394&3 & 700&7\\
100 -- 120 & 52859 & 52920 & Aug.&8,  &2003 & Oct. &8,  &2003 &
  717 & 1843&6 & 709&0\\
164 -- 185 & 53051 & 53115 & Feb.&16, &2004 & Apr. &20, &2004 &
  592 & 1317&0 & 519&0\\
222 -- 246 & 53225 & 53298 & Aug.&8,  &2004 & Oct. &20, &2004 &
  698 & 1590&3 & 618&7\\
283 -- 307 & 53407 & 53481 & Feb.&6,  &2005 & Apr. &21, &2005 &
 1027 & 2256&9 & 1116&4\\
345 -- 366 & 53594 & 53655 & Aug.&12, &2005 & Oct. &12, &2005 &  
  470 & 1009&6 & 450&5\\
399 -- 431 & 53756 & 53850 & Jan.&21, &2006 & Apr. &25, &2006 &
  517 & 1405&8 & 534&0\\
466 -- 485 & 53960 & 54013 & Aug.&13, &2006 & Oct. &5,  &2006 &
  434 & 1141&6 & 542&6\\

\hline
\end{tabular}

\label{tab:rxsobs}
\end{table*}

The {\em INTErnational Gamma-Ray Astrophysics Laboratory} (INTEGRAL)
is a hard X-ray/soft gamma-ray (3~keV -- 8~MeV) mission which has been
operational since October 2002. The payload consists of two main
coded-mask imaging telescopes. 1) IBIS, {\em Imager on Board the
  INTEGRAL Satellite} \citep{Ubertini03_ibis}, has a wide field of
view (FOV) of $29^\circ \times 29^\circ$ (full-width zero response)
and it has an $\sim$12\arcmin\ angular resolution.  The low-energy
detector is called the {\em INTEGRAL Soft Gamma-Ray Imager}
\citep[ISGRI;][]{Lebrun03_ISGRI} and it is sensitive between
$\sim$20~keV and $\sim$300~keV. 2) The {\em Spectrometer for INTEGRAL}
\citep[SPI; ][]{Vedrenne03_SPI} has a better spectral resolution but
only a moderate imaging capability with an angular resolution of
$\sim$2\fdg5. SPI is sensitive between $\sim$20~keV and $\sim$8~MeV
and has a FOV of $35^\circ \times 35^\circ$ (full-width zero
response).

For the best sky map reconstruction INTEGRAL performs observations in
dither patterns \citep{Jensen03_igr} besides the occasional staring
observations. Typical pointings (Science Windows, ScWs) last
$\sim$1800--3600~s. Due to the limited visibility windows on the
Galactic-centre region, the observations form sets of consecutive
orbital revolutions (Revs). In Table~\ref{tab:rxsobs} a list of these
sets are given with the total exposure time for the selected ScWs
for which \rxs1708 was within an angle of 14\fdg5 from the pointing
direction of IBIS, as well as the effective on-source exposure, which
is reduced due to off-axis viewing angles.

The region where \rxs1708 is located is often observed with \igr\ as
it is $\sim$15\degr\ from the Galactic centre. After four years of
observations the total exposure on this region adds up to 12~Ms taken
in 5161 pointings, after screening for Solar flares and erratic
count-rates due to the passages of the spacecraft through the Earth's
radiation belts. Taking into account the off-axis angles the effective
on-source exposure is 5.2~Ms.

\subsubsection{INTEGRAL spectra from spatial analysis}
\label{sec:rxsigrspec}

The spectral analysis of the IBIS-ISGRI data follows the same
procedures as applied in \citet{denHartog08_0142}. The shadowgrams of
5161 ScWs are deconvolved into sky images in 20 energy intervals using
the Off-line Scientific Analysis software package version 5.1
\citep[OSA; see][]{Goldwurm03_osa}. The energy intervals are
exponentially binned between 20~keV and 300~keV. At the source
position of \rxs1708 averaged count rates are determined for all
energy bins. Photon fluxes are generated by normalizing the AXP
count-rates to the Crab count-rates determined from \igr\ Crab
observations during Rev.~102, and the Crab spectrum. The spectral
shape of the Crab spectrum we adopted is a curved power-law shape
\begin{equation}
\label{eq:rxscrab}
F_{\gamma} = 1.5703(14)
\times (E_{\gamma}/0.06335)^{-2.097(2)-0.0082(16)\times {\rm
ln}(E_{\gamma}/0.06335)} 
\end{equation}
\citep[Eq.~2 in][]{Kuiper06_axps}, where $F_{\gamma}$ is expressed in
ph/(cm$^2$\,s\,MeV) and $E_{\gamma}$ in MeV. For fitting these (and
other) spectra the software package {\it X-spec} version 12.3
\citep{Arnaud96_xspec} and private software are used. All errors
quoted in this paper are 1$\sigma$ errors.

For the \igr\ observations up to Rev.~366 SPI spectral analysis is
performed above 140~keV, from where the sensitivity of SPI is
comparable or higher than ISGRI. In the 140~keV -- 1~MeV energy range
the data were binned in 4 energy bins. From the total-exposure sky maps
source fluxes have been extracted for every energy bin using a maximum
likelihood fitting procedure that considers the instrumental response
of SPI \citep{Knoedlseder04_spi}. Also here the measured flux is
normalized to the Crab spectrum, which has been obtained using SPI
observations of Revs. 43--45.

\subsubsection{INTEGRAL timing analyses}
\label{sec:rxsigrtiming}
\begin{table*}
\centering
\renewcommand{\tabcolsep}{1.7mm}
\caption[]{Phase-coherent ephemerides -- derived from RXTE-PCA
monitoring data -- valid for the analysed INTEGRAL observations.  }

\begin{tabular}{l l l l l l l r@{.}l l}

\vspace{-3mm}\\

\hline
\hline
\vspace{-3mm}\\
AXP &
Start &
End &
INTEGRAL &
$t_{\rm Epoch}$ &
$\nu$ &
$\dot{\nu}$ &
 \multicolumn{2}{l}{$\ddot{\nu}$} &
$\Phi_0$ \\

 & [MJD] & [MJD] & range (Revs) & [MJD, TDB] &[Hz] & 
 $\times 10^{-13}\,\rm{[Hz\,s^{-1}]}$&  
 \multicolumn{2}{l}{$\times 10^{-22}\,\rm{[Hz\,s^{-2}]}$}& \\
\hline
\vspace{-3mm}\\
\rxs1708 & 52590 & 52960 & 009--133 & 52590 & 0.0908982198657
 & $-$1.58268 & 0&00 & 0.4333\\
\rxs1708 & 53050 & 53325 & 163--255 & 53050 & 0.0908919724436
 & $-$1.62994 & 2&75 & 0.5025\\
\rxs1708 & 53321 & 53520 & 254--320 & 53377 & 0.0908875116829
 & $-$1.49563 & $-$34&1 & 0.7559\\
\rxs1708 & 53555 & 53690 & 332--377 & 53555 & 0.0908852012859
 & $-$1.77561 & 36&4 & 0.8587\\
\rxs1708 & 53741 & 54055 & 395--499 & 53741 & 0.0908825979045
 & $-$1.57228 & $-$0&813 & 0.1842\\

\hline
\end{tabular}

\label{tab:rxseph}
\end{table*}

\begin{table}
\centering
\renewcommand{\tabcolsep}{1.3mm}
\caption[]{Summary of the RXTE observations of \rxs1708. The following
  information is given: the observation IDs, time spans of the
  observations in calendar dates, PCA exposure times for PCU-2 in
  units ks.}

\begin{tabular}{l l@{ }r l@{ -- } l@{ }l l r@{.}l}

\vspace{-3mm}\\

\hline
\hline
\vspace{-3mm}\\
Obs. ID. &
\multicolumn{6}{c}{Time span} &

\multicolumn{2}{c}{$t_{\mathrm{PCA}}$}\\
\hline
\vspace{-3mm}\\
ALL   & Jan.&12, &1998 & Nov. &16, &2006 & 606&512 \\
Set A & Jan.&12, &1998 & Oct. &26, &2003 & 310&480 \\
Set B & Nov.& 1, &2003 & Nov. &16, &2006 & 296&032 \\
\hline
\vspace{-3mm}\\
\multicolumn{8}{c}{{\bf Set A}}\\
30125 & Jan.&12, &1998 & Jan. &08, &1999 & 59&896 \\
40083 & Feb.&6,  &1999 & Mar. &11, &2000 & 52&568 \\
50082 & Apr.&21, &2000 & May  &12, &2001 & 34&576 \\
60069 & May &6,  &2001 & Feb. &20, &2002 & 24&544 \\
60412 & May &20, &2001 & May. &23, &2001 &  9&928 \\
70094 & Apr.&2,  &2002 & Mar. &20, &2003 & 55&600 \\
80098 & Apr.&16, &2003 & Oct. &26, &2003 & 73&368 \\
\multicolumn{8}{c}{{\bf Set B}}\\
80098 & Nov.&1,  &2003 & Feb. &23, &2004 & 26&416 \\
90076 & Feb.&28, &2004 & Feb. &28, &2005 &126&784 \\
91070 & Mar.&6,  &2005 & Feb. &25, &2006 & 79&000 \\
92006 & Mar.&5,  &2006 & Nov. &16, &2006 & 63&832 \\
\hline
\end{tabular}

\label{tab:rxsrxteobs}
\end{table}

For the IBIS-ISGRI timing analysis of AXP~\rxs1708 we followed earlier
applied procedures \citep{Kuiper06_axps, denHartog08_0142}.  All
available data taken between Rev.~037 and 485 were used (see
Table~\ref{tab:rxsobs}). Photons were selected only from non-noisy
detector elements that could be illuminated by the source through the
open mask elements by more than 25\%. After correcting for
instrumental, onboard processing and ground-station time delays
\citep{Walter03_timing}, the event arrival times are Solar-system
barycentered (JPL DE200 Solar-system ephemeris) adopting the best
known source position \citep{Israel03_1708}.  The barycentered events
are folded using a phase-connected ephemeris (see
Table~\ref{tab:rxseph} and Sect.~\ref{sec:rxsrxte} for the ephemeris
creation). The (TDB) time to pulse phase conversion taking into
account consistent phase alignment for each ephemeris is provided by
the following formula:
\begin{equation}
\Phi(t) = \nu \cdot (t-t_{\rm Epoch}) + \frac{1}{2}\dot{\nu}\cdot
(t-t_{\rm Epoch})^2 + \frac{1}{6}\ddot{\nu}\cdot (t-t_{\rm Epoch})^3 -
\Phi_0.
\end{equation}

For the pulse profiles, detection significances are estimated by
applying a Z$^2_{{\rm n}}$ test on bin-free pulse-phase distributions.
\citep{Buccheri83_zn2}. Truncated Fourier series with two
harmonics are used to fit the pulse profiles, where the minimum of
these fits are defined as the DC level.

The total-pulsed spectrum is derived by obtaining the number of excess
counts above the DC level from each pulse profile for all considered
energy bands. These excess counts are normalized in Crab units (by
applying the same procedure for the Crab pulsar) and are converted
into photon fluxes using a Crab-pulsar spectrum model \citep[Eq.~3
  in][]{Kuiper06_axps};
\begin{equation}
F_{\gamma} = 0.4693(21)
\times (E_{\gamma}/0.04844)^{-1.955(7)-0.0710(78)\times {\rm
ln}(E_{\gamma}/0.04844)}.
\end{equation}

The analysis to create phase-resolved spectra
(Sect.~\ref{sec:rxsphaseres}) is similar. However, for the
phase-resolved spectra only the excess counts within selected phase
intervals are used.


\subsection{RXTE}
\label{sec:rxsrxte}

For the timing analysis of the \igr\ data the regular monitoring
observations with the {\em Rossi X-ray Timing Explorer} (RXTE) during
the INTEGRAL observations are of great importance
\citep{Gavriil02_rxtemonitoring}, because the \igr\ source count rate
is too low to independently detect the pulsation. This is possible
with the data from the Proportional Counter Array
\citep[PCA;][]{Jahoda96_pca} aboard RXTE. The PCA is a non-imaging
instrument sensitive in the 2--60 keV energy range and it consists of
five collimated units (PCUs 0--4) with a $\sim$1\degr\ FOV. Using
these \pca\ monitoring data we created accurate phase-connected
ephemerides valid during the \igr\ observations.  The details of the
ephemerides can be found in Table~\ref{tab:rxseph}. The pulsar phase
was set arbitrarily to zero where the pulse profile for energies
between $\sim$2.5 and 10 keV reached a sharp minimum (see
e.g. Fig.~\ref{fig:rxsrxte8}\,B). This is a different choice than made
in \citet{Kuiper06_axps}, resulting in a phase shift in the profiles
of $\sim$0.2.

For the study of the pulsed emission of \rxs1708 with \pca, we have
used all publicly available data spanning nine years of
observations. The data are analysed in two sets. Set~A contains the
data which have been used by \citet{Kuiper06_axps} and were taken over
a period of almost six years. Most of these data were taken before the
start of the \igr\ observations. Set~B covers the remaining part of
the publicly available data taken during \igr\ operations and span
about three years. All data used in this work have been listed in
Table~\ref{tab:rxsrxteobs}. It can be seen that Set A and B have
similar total exposures on \rxs1708.

The procedures for the RXTE spectral-timing analysis as outlined by
\citet{Kuiper06_axps}, and also applied in \citet{denHartog08_0142},
have been followed.  The pulse profiles for any selected
energy band can be described sufficiently accurate by truncated
Fourier series with three harmonics above constant DC levels. The
excess (pulsed) counts above these DC levels can then be
converted into flux units using PCU exposure-weighted response
matrices \citep[see e.g. Sect.~3.2 of][]{Kuiper06_axps}. To derive
unabsorbed PCA time-averaged total-pulsed and phase-resolved spectra,
a Galactic absorption column $N_{{\rm H}} = 1.47 \times 10^{22}$
cm$^{-2}$ was used (fixed value), derived
from \xmm\ spectral fits (see Sect.~\ref{sec:rxsxmmtot}).

The High-Energy X-Ray Timing Experiment \citep[HEXTE; 15-250
  keV;][]{Rothschild98_hexte} aboard RXTE has also been used to extend
the PCA spectral-timing study to higher energies. HEXTE consists of
two independent detector clusters (0 and 1) each with a
$\sim$1\degr\ FOV.  HEXTE allows for detailed consistency checks with
the \igr\ results obtained for similar energy windows. Due to the
co-alignment of HEXTE and the PCA, the same long monitoring
observations of \rxs1708 can be used for both instruments. Fully
consistent procedures have been applied for the spectral-timing
analysis, including the use of the ephemerides listed in
Table~\ref{tab:rxseph}, and following the same procedures for deriving
pulsed counts and fluxes of the pulsed emission. On \rxs1708 the total
deadtime corrected exposures for clusters 0 and 1 are 239.04 ks and
259.50 ks, respectively.


\subsection{XMM-Newton}
\label{sec:rxsxmm}

\xmm\ \citep{Jansen01_xmm} has been operational since early
2000. Onboard are three CCD cameras for X-ray imaging, namely two EPIC
(European Photon Imaging Camera) MOS \citep{Turner01_mos} cameras and
one EPIC-PN \citep{Strueder01_pn} camera. All cameras have a FOV of
$\sim$30\arcmin\ and are sensitive in the energy range $\sim$0.3--12
keV.

We have (re)analysed a publicly available 45~ks \xmm\ observation on
\rxs1708. The data were taken in the night of August 28 to 29, 2003
(Obs. Id 0148690101). The data are analysed using SAS v.~7.0 and the
latest calibration files that were available (May 2007). Next, we
considered the EPIC-PN which operated in small-window mode with medium
filter. In this mode the maximum time resolution is 6 ms. These data
are checked for solar (soft proton) flares by creating a light curve
with events with energies larger than 10~keV.  A count-rate
distribution was created from this light curve to which a Gaussian was
fitted to determine the mean count rate of the high-energy
photons. Good-Time Intervals (GTIs) were created allowing only time
stamps for which the count rate was lower than 0.078 counts per second
which corresponds to the fitted mean count rate plus three times the
width of the distribution (3$\sigma$). The GTIs add up to 25.5 ks of
exposure. All single and double events, i.e. patterns less than and
equal four, are selected within the energy range 0.3~keV to
12~keV. The arrival times of the selected events are barycentered.

There was no need to correct for pile-up effects as the count rate of
\rxs1708\ was low enough for the PN in small window mode with medium
filter. To extract the source counts a circular extraction region with
a radius of 35\arcsec\ was used. As this source is not so bright it
was possible to select a background extraction region from the small
window with a radius of 45\arcsec. No out-of-time event correction has
been applied as the effect is approximately 1\%.  The extracted
spectrum has been binned oversampling the energy resolution by a
factor of three and then rebinned occasionally (for energies
$\gtrsim$5 keV) to ensure a minimum of 25 counts per bin.

For timing analysis, the barycentered events are folded using an
appropriate ephemeris (ephemeris 1 in Table~\ref{tab:rxseph}).  To
create pulsed spectra pulsed excess counts are extracted as described
in Sect.~\ref{sec:rxsigrtiming}, fitting the pulse profiles with
truncated Fourier series with the first three harmonics above a
constant background.

\begin{table}[t]
\centering
\renewcommand{\tabcolsep}{1.7mm}
\caption[]{Summary of the power-law model fits to the
  INTEGRAL-IBIS-ISGRI total-spectra of \rxs1708 summing observations
  taken in intervals of $\sim$1 year and for the total exposure.  For
  each spectrum are given the revolution interval over which \rxs1708
  observations are summed, the photon index, the 20--150 keV flux and
  \chir.}

\begin{tabular}{l r@{ $\pm$ }l r@{ $\pm$ }l l }

\vspace{-3mm}\\

\hline
\hline
Rev.&
\multicolumn{2}{c}{$\Gamma$}& 
\multicolumn{2}{c}{$F_{20-150} \times 10^{-11}$} & 
\chir (dof) \\

 &  \multicolumn{2}{c}{}  & \multicolumn{2}{c}{[\ecs]}&\\
\hline

037--120 & 1.18&0.14 & 5.47&0.43 & 1.64 (14) \\
164--246 & 1.40&0.15 & 5.97&0.55 & 1.09 (14) \\
283--366 & 0.97&0.12 & 7.13&0.44 & 0.52 (14) \\
399--485 & 1.00&0.21 & 8.72&0.61 & 1.67 (14) \\
\hline
037--485 & 1.13&0.06 &  6.61&0.23$^\dagger$ & 1.13 (14)\\
\multicolumn{6}{l}{$^\dagger F_{20-175} = (7.76 \pm 0.34) \times 10^{-11}$ \ecs}\\
\hline
\end{tabular}

\label{tab:rxsfits}
\end{table}


\section{Results}
\label{sec:rxsresults}

In this section we present results from \igr, \xmm\ and RXTE. First,
we present the total (pulsed plus DC) spectrum using \igr\ data above
20~keV and look for possible long-term time variability. Then we
derive the complementary \xmm\ total spectrum below 12~keV. RXTE is a
non-imaging mission and due to the presence of nearby strong sources
and large gradients in the Galactic ridge emission
\citep{Valinia98_ridge}, it can in this work only be used for studying
the timing signal of \rxs1708. The pulse profiles are then presented
and compared for all three missions. Finally, the results from
spectral timing analyses in the form of total-pulsed spectra and
phase-resolved spectra for all data sets are presented and discussed
in detail. This leads to the identification of genuinely different
components contributing to the pulsed high-energy emission.


\subsection{Total spectra and long-term variability}
\label{sec:rxstotspec}
\begin{figure}
\psfig{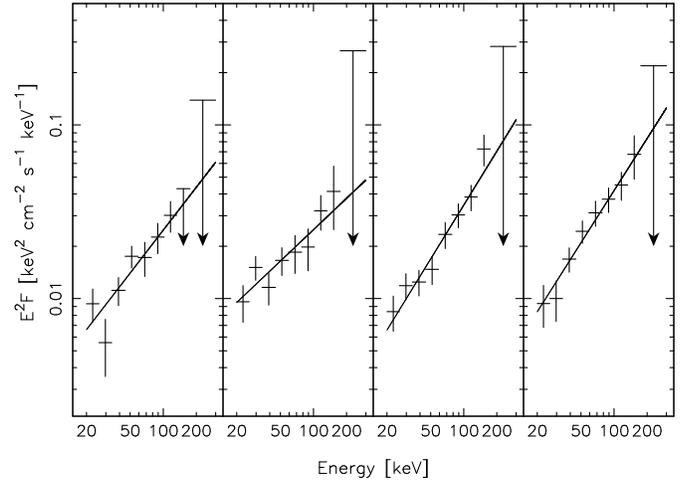}
\caption{Total-flux (pulsed + DC) spectra of \rxs1708 collected during
  four one-year intervals of INTEGRAL observations, corresponding to
  revolution intervals (from left to right); 037--120, 164--246,
  283--366 and 399-485. The upper limits are 2$\sigma$. The model
  lines are single power-law fits (see Table~\ref{tab:rxsfits}).
\label{fig:rxs4spec}}
\end{figure}

\subsubsection{INTEGRAL ISGRI persistent time-averaged spectra 
and long-term variability}
\label{sec:rxsigrvar}

\begin{figure}
\psfig{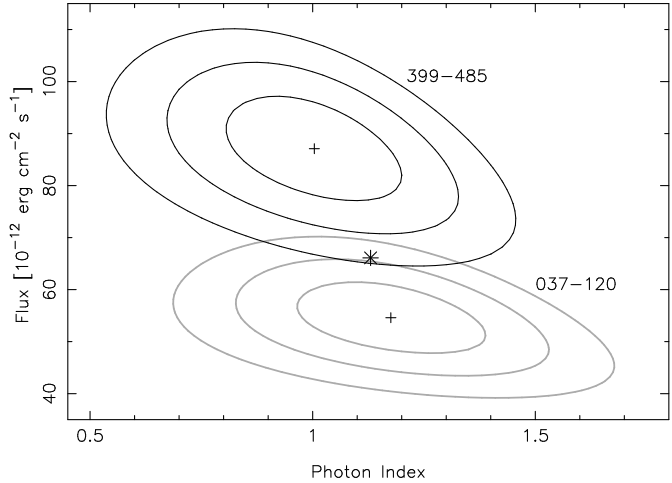}
\caption{Error contours ($1\sigma, 2\sigma$ and $3\sigma$) of the
  power-law model fits to the total spectrum (20--300 keV) for
  observations 037-120 and 399-485 compared to the total time-averaged
  fit (037-485).  Out of the four observations (each averaged over one
  year) these two differ most from the four-year time-averaged
  spectrum.
\label{fig:rxscontours}}
\end{figure}

Listed in Table~\ref{tab:rxsobs} are eight sets of \igr\ observations,
each covering about half a year. \rxs1708 is not bright in hard
X-rays, therefore we first performed the spectral analysis averaging
observations over one-year time intervals to obtain significant flux
values till above 100 keV.  The resulting four spectra are shown in
Fig.~\ref{fig:rxs4spec}. They can be described satisfactorily with
single power-law models. The fit results are listed in
Table~\ref{tab:rxsfits}. The power-law indices range from $0.97 \pm
0.12$ to $1.40 \pm 0.15$ and the 20--150~keV fluxes range from $(5.47
\pm 0.43) \times 10^{-11}$ to $(8.72 \pm 0.61) \times 10^{-11}$
\ecs\ which can be seen as a indication for time variability. When we
compare these yearly averages with the long-term four year average
(see Sect.~\ref{sec:rxsigrtot}), we find that the spectra taken in
Revs. 037--120 and 399--485 are most deviating from this time-averaged
spectrum, i.e. at the $\sim$2$\sigma$ and $\sim$3$\sigma$ level,
respectively (see error contours in Fig.~\ref{fig:rxscontours}).

Alternatively, the standard deviations ($s = \sqrt{ 1/(n-1) \sum_i
  (x_i - \bar{x})^2 }$) for the power-law indices and the 20--150 keV
fluxes relative to the weighted mean are 0.20 and $1.46 \times
10^{-11}$ \ecs, respectively. Therefore, the power-law shape is stable
within 18\% (1$\sigma$) and the 20--150~keV flux is stable within 22\%
(1$\sigma$).

\begin{figure}
\psfig{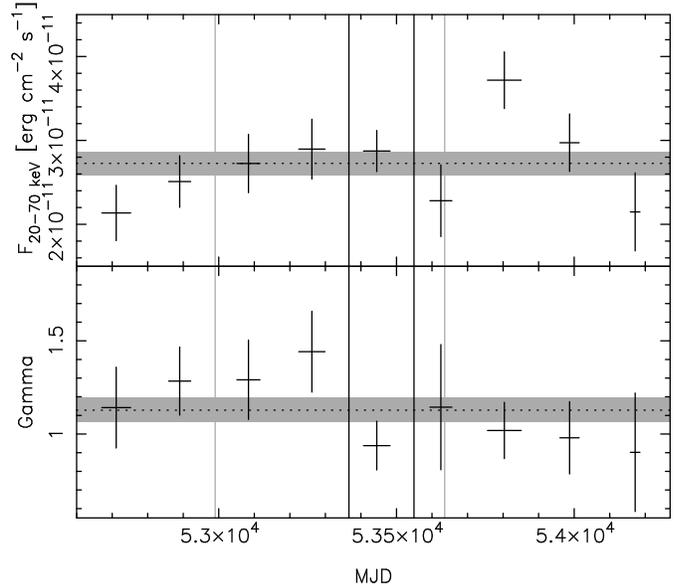}
\caption{Fluxes (20--70 keV) and photon indices for the half-year time
  sets specified in Table~\ref{tab:rxsobs} and one shorter observation
  (see Sect.~\ref{sec:rxsigrvar}). Indicated as horizontal dotted
  lines are the time-averaged values with their 1$\sigma$ errors (grey
  bands). Glitches and candidate glitches
  \citep{Israel07_rxs1708,Dib08_glitches} are indicated as vertical black
  and grey lines, respectively.
\label{fig:rxsfluxes}}
\end{figure}

To explore the indication for variability further, we also fitted the
spectra of the half-year long time sets (Table~\ref{tab:rxsobs}) plus
a more recent short, lower-statistics observation (Revs 534--542,
Feb. 27 -- Mar. 25, 2007, 173 ScWs, 307 ks effective on-source
exposure) with single power-law models, but over the narrower energy
band 20--70~keV.  The results are plotted in
Fig.~\ref{fig:rxsfluxes}. The 20--70~keV flux measurement for
Revs.~399--431 is the only one that deviates at a 3$\sigma$ level from
the time-average value. Since this is one single 3$\sigma$ deviation
for nine measurements, we consider this only as an indication for time
variability.

The bottom panel of Fig.~\ref{fig:rxsfluxes} shows no indication for
significant variation in spectral index, either.
\citet{Israel07_rxs1708} and \citet{Dib08_glitches} reported the detection
of glitches and candidate glitches of \rxs1708 during the INTEGRAL
operations, the epochs are indicated in Fig.~\ref{fig:rxsfluxes}. This
figure shows that there is no indication for an obvious change in
spectral slope nor flux triggered by one of the glitches. Possibly
there is a change in spectral slope after the glitch on MJD
53366. However, the difference between $\Gamma = 1.27 \pm 0.10$ and
$\Gamma = 0.98 \pm 0.08$, the average values before and after the
glitch, is not significant (2.3$\sigma$).

\citet{Gotz07_1708} analysed part of the \igr\ data used in this work,
they did not use Revs. 164--185, 345--366 and 399--431, and claim the
detection of long-term hard X-ray (20--70 keV) variability based on an
indication for lower flux values in the \igr\ measurements after the
second candidate glitch (epoch marked in Fig.~\ref{fig:rxsfluxes}),
correlated with lower fluxes measured with {\it Swift} in the 1--10
keV band (their Fig.~1).  However, our more complete analysis
including full spectral fitting of the data in stead of using merely
count rates does not confirm the reported trend in the \igr\ data for
\rxs1708 (see Fig.~\ref{fig:rxsfluxes}). Therefore, we do not support
the claimed correlated variability between the hard X-ray and soft
X-ray emission from \rxs1708.

\subsubsection{INTEGRAL time-averaged total spectrum}
\label{sec:rxsigrtot}

\begin{figure}
\psfig{figure=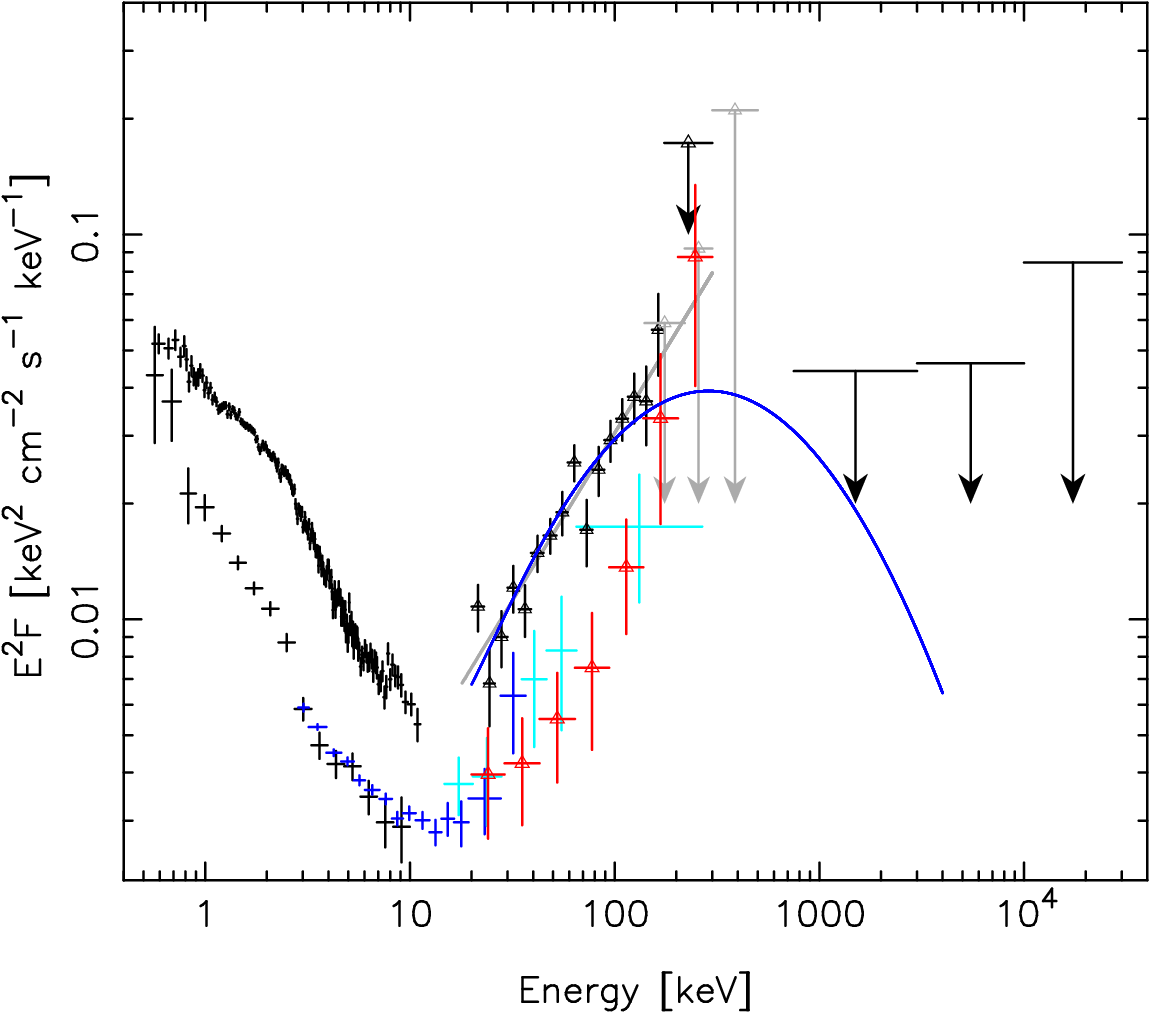,width=\columnwidth,angle=-90,clip=t}
\caption{High-energy spectra of \rxs1708. In this figure the following
  is plotted: the unabsorbed total spectra of \xmm\ ($<$12~keV) and
  \igr\ (with triangle markers) in black, also in black three COMPTEL
  upper limits \citep{Kuiper06_axps}, three \igr-SPI upper limits in
  grey (with triangle markers); also in grey a power-law fit to the
  \igr-IBIS spectrum, in blue a logparabolic fit to the \igr-IBIS, SPI
  and COMPTEL data, total pulsed spectra of \xmm\ in black, \pca\ and
  HEXTE are shown in blue and aqua, and the total pulsed spectrum of
  \igr-ISGRI in red (with triangle markers).
\label{fig:rxstothigh}}
\end{figure}

Combining all the data listed in Table~\ref{tab:rxsobs} a
time-averaged spectrum is extracted using 5.2~Ms on-source exposure
time (Fig.~\ref{fig:rxstothigh}). \rxs1708 has been detected with high
significance up to 175~keV with a 4.2$\sigma$ detection significance
in the 152--175~keV band. Fitting this spectrum with a single power
law yields an excellent fit (\chir = 1.13 with 14 dof) with a photon
index $\Gamma = 1.13 \pm 0.06$ and a 20--150~keV flux of $(6.61 \pm
0.23) \times 10^{-11}$ \ecs. Its 20--175~keV luminosity is $L_{20-175}
= 1.34 \times 10^{35}$ \es\ assuming a distance of 3.8 kpc
\citep{Durant06_distances}. It is remarkable that there is no hint for
a spectral break up to the highest energies in our INTEGRAL-ISGRI
spectrum, particularly when compared to the reported COMPTEL upper
limits for energies above 750 keV \citep{Kuiper06_axps}, which impose
a spectral break between 100 keV and 750 keV (see
Fig.~\ref{fig:rxstothigh}).

Unfortunately, it turned out that \rxs1708\ is not bright enough above
$\sim$100~keV to be detected with INTEGRAL-SPI, even for this long
exposure. In Fig.~\ref{fig:rxstothigh} three 2$\sigma$ upper limits
derived from SPI spatial analysis are added.  These limits are all
above the extrapolation of the power-law fit to the ISGRI flux values,
contrary to the case of \0142, for which SPI upper limits provided
evidence for the presence of a spectral break
\citep{denHartog08_0142}.

Following \citet{denHartog08_0142} we have fitted all \igr\ and
COMPTEL spectral information (including limits) with a logparabolic
function;
\begin{equation}
\label{eq:rxslogpar}
F = F_{0} \times \left(\frac{E}{E_{0}}\right)^{-\alpha -\beta
  \cdot{\rm ln}\left(\frac{E}{E_0}\right)}
\end{equation}
where $E_0$ (in units keV) is the pivot energy to minimize
correlations between the parameters and $F_0$ is the flux (in units
\pcsk) at $E_0$. This function is a power-law if the curvature
parameter $\beta$ is equal to zero.  Assuming this spectral shape we
get an acceptable broad-band (20 keV -- 30 MeV) fit with best-fit
parameters $\alpha = 1.637 \pm 0.049, \beta = 0.261 \pm 0.035$ and
$F_0 = (1.68 \pm 0.08) \times 10^{-6}$ \pcsk\ at $E_0 = 143.276$
keV. The peak energy $E_{{\rm peak}}$ is $ 287^{+75}_{-45}$ keV. This
value lies remarkably close to the peak energy found for
\0142\ \citep[i.e. $279^{+65}_{-41}$ keV;][]{denHartog08_0142}.  In
Fig.~\ref{fig:rxstothigh} both the power-law and the logparabolic fit
are drawn.

\subsubsection{\xmm\ total spectrum}
\label{sec:rxsxmmtot}

For energies below 12~keV we extracted the absorbed total (pulsed +
DC) spectrum using \xmm\ EPIC-PN data (see Sect.~\ref{sec:rxsxmm}). In
order to obtain an estimate for the Galactic absorption column
($N_{{\rm H}}$) we fitted the spectrum globally with a canonical
logparabolic function, including fixed \igr\ parameters for the hard
X-ray contribution above $\sim$8~keV. We derive an $N_{{\rm H}}$ of
$(1.47 \pm 0.02) \times 10^{22}$ cm$^{-2}$, which can be compared with
the value ($1.36 \pm 0.03) \times 10^{22}$ cm$^{-2}$ obtained by
\citet{Rea05_1708} fitting the same \xmm\ data with an absorbed
black-body plus a power-law model.  \citet{Durant06_extinction} used a
model-independent approach analysing X-ray grating spectra taken with
the Reflection Grating Spectrometer \citep{denHerder01_rgs} onboard
\xmm. Their value for $N_{{\rm H}}$ of $(1.40 \pm 0.4)\times 10^{22}$
cm$^{-2}$ is consistent with both estimates. We adopted $N_{{\rm H}} =
1.47 \times 10^{22}$ cm$^{-2}$ in this work for the
\xmm\ and RXTE analyses. The total unabsorbed spectrum is shown in
Fig.~\ref{fig:rxstothigh}. The 2--10~keV unabsorbed flux is $(3.398
\pm 0.012) \times 10^{-11}$ \ecs. The error is statistical only.  The
2--10~keV unabsorbed fluxes for $N_{{\rm H}} = 1.40 \times 10^{22} $
and $1.36 \times 10^{22}$\ cm$^{-2}$ are $(3.361 \pm 0.009) \times
10^{-11}$ and $(3.339 \pm 0.013) \times 10^{-11}$ \ecs,
respectively. These values are within 2\% of our value.


\subsection{Pulse profiles}
\label{sec:rxspp}
\begin{figure}
\psfig{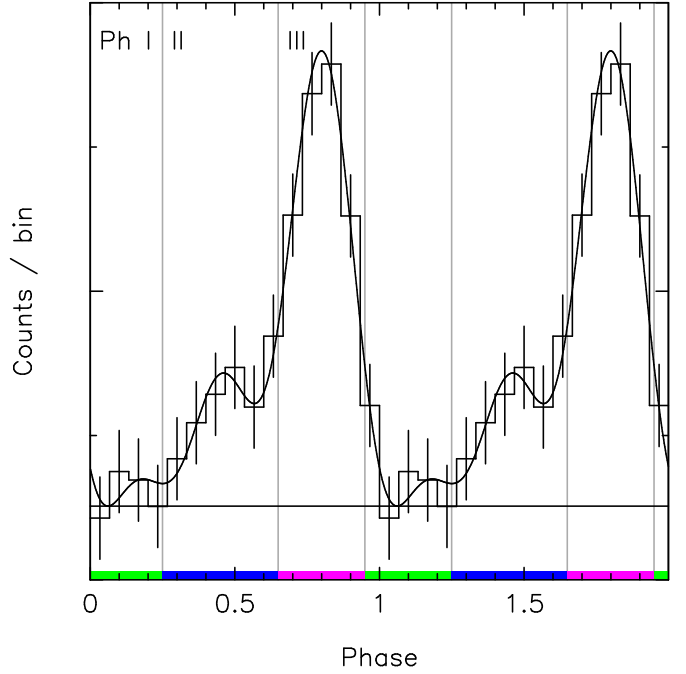}
\caption{IBIS-ISGRI pulse profile of \rxs1708 between 20 keV and 270
  keV.  This profile has a 12.3$\sigma$ significance using a Z$^2_3$
  test (fit shown as a solid curve). The fitted DC level is indicated
  with an horizontal line. The grey lines and the colours indicate
  three phase intervals Ph\,I, II and III (see
  Table~\ref{tab:rxsph}). The colours are consistently used in this
  paper in figures showing results of phase-resolved analyses (see
  Sect.~\ref{sec:rxsphaseres}).
\label{fig:rxstotpuls}}
\end{figure}
\begin{figure}
\centering
\psfig{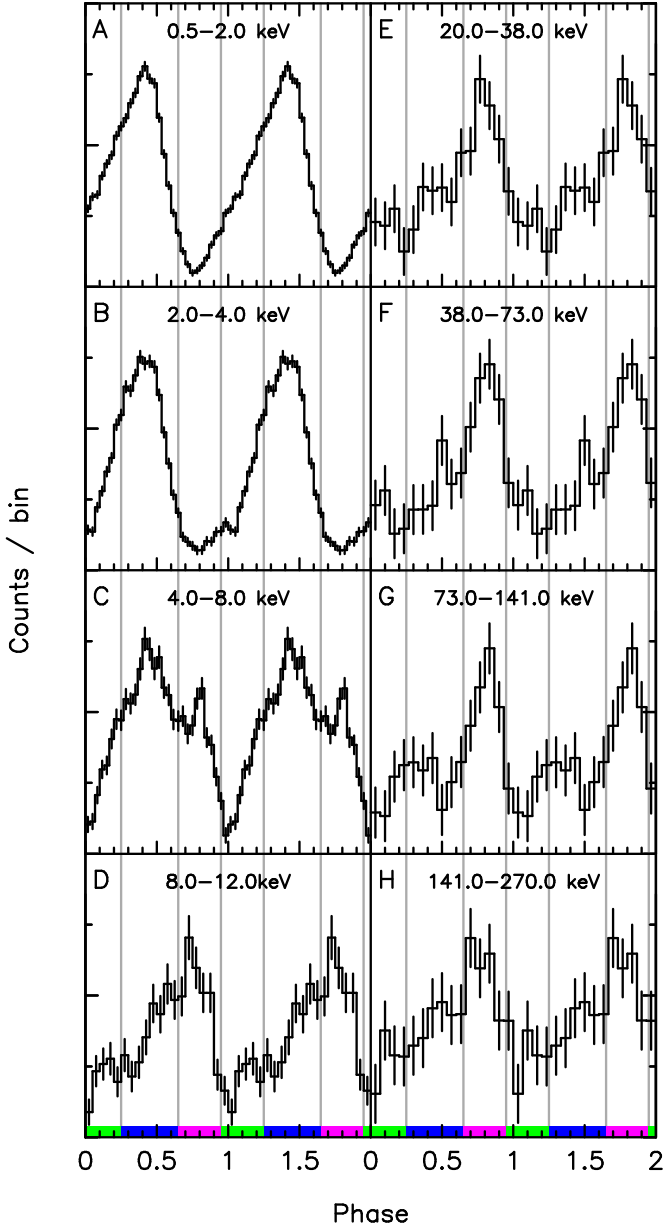}
\caption{Pulse profiles from soft to hard X-rays. \xmm\, pulse
  profiles in the energy range 0.5--12.0~keV are shown in panels A, B,
  C and D.  Panels E, F, G and H show \igr\, pulse profiles in the
  energy range 20--270~keV. The differential energy ranges are
  indicated in the figures. The phase intervals are indicated as in
  Fig.~\ref{fig:rxstotpuls}.
\label{fig:rxs8panel}}
\end{figure}
\begin{figure}
\centering
\psfig{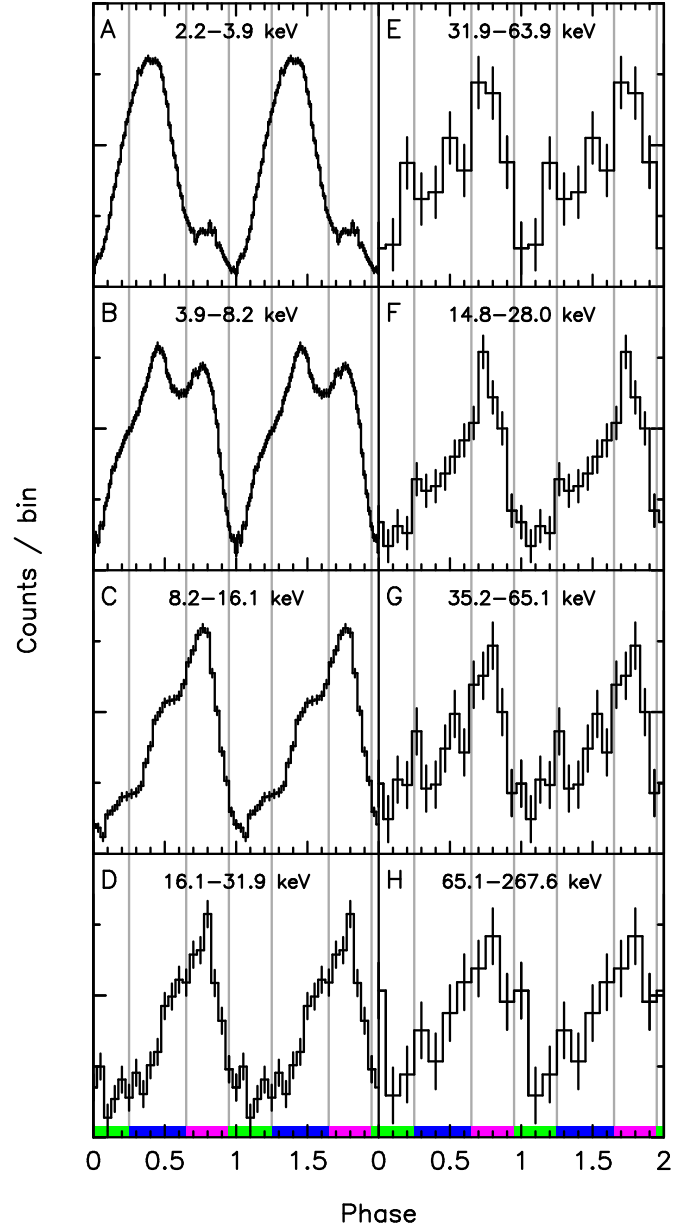}
\caption{Pulse profiles from RXTE.  \pca\ pulse profiles in the energy
  range 2.2--63.9~keV are shown in panels A, B, C, D and E.
  RXTE-HEXTE pulse profiles are shown in panels F, G and H within the
  energy range 14.8--267.6~keV. The phase intervals are indicated as
  in Fig.~\ref{fig:rxstotpuls}.
\label{fig:rxsrxte8}}
\end{figure}

\subsubsection{\igr\ and \xmm\ pulse profiles}
\label{sec:rxsigrxmmpp}

\citet{Kuiper06_axps} showed for the first time pulsed hard X-ray
emission ($>$10~keV) from \rxs1708 using data from \pca, RXTE-HEXTE
and \igr-ISGRI. For the \igr\ pulse profiles $\sim$1.4~Ms on-source
exposure was used, resulting in a 5.9$\sigma$ detection for energies
20--300~keV.  In this work, we present \igr\ pulse profiles using
$\sim$5.2~Ms on-source exposure.  The result is a very much improved
pulse profile with a 12.3$\sigma$ detection significance
\citep[Z$^2_{{\rm 3}}$ test;][]{Buccheri83_zn2} for
energies 20--270~keV (Fig.~\ref{fig:rxstotpuls}). The profile shows a
single pulse which peaks around phase 0.8 with a steep trailing wing
dropping off to the DC level at phase $\sim$1.05 (0.05). On the
leading wing there appears to be a weak pulse or shoulder.

Presented in Fig.~\ref{fig:rxs8panel}\,E--H are four
exponentially-binned differential \igr\ pulse profiles. The profiles
have significances of 6.3$\sigma$, 6.8$\sigma$, 6.2$\sigma$ and
3.5$\sigma$, respectively, using a Z$^2_{{\rm 2}}$ test. All profiles
show single pulses which all peak at phase $\sim$0.8, like for the
total pulse in Fig.~\ref{fig:rxstotpuls}.

The complementary high-statistics \xmm\ pulse profiles at lower
energies (0.5--12~keV) are shown in
Fig.~\ref{fig:rxs8panel}\,A--D. The pulse profile with the lowest
statistics, in panel D, still has a significance of 10.7$\sigma$
(Z$^2_{{\rm 3}}$ test). The drastic change in morphology moving up in
energy from 0.5 keV up to 270 keV is evident and will be investigated
further. Note that in the timing analysis the pulsed emission has been
detected to higher energies than the total emission in the sky maps.

\subsubsection{RXTE-PCA and RXTE-HEXTE pulse profiles}
\label{sec:rxsrxtepp}

The \pca\ bridges the observational gap (12--20 keV) in energy between
\xmm\ and \igr, with \xmm\ being sensitive down to $\sim$0.5 keV and
\igr\ extending the coverage to $\sim$300 keV. In addition, RXTE-HEXTE
is sensitive over about the same energy band as INTEGRAL-ISGRI.
Analysis of data from these three missions allow for consistency
checks, particularly when the observations are covering the same
epochs and for checks on long-term variability.  RXTE observations
set~A (Table~\ref{tab:rxsrxteobs}) covers about 6 years (mostly before
the launch of INTEGRAL) and was used by \citet{Kuiper06_axps}.  Set~B
overlaps in time with the INTEGRAL observations up to December
2006. Therefore, we first compared the time-averaged \pca\ pulse
profiles in 5 differential energy bands for set~A with those for
set~B. We did not find significant differences in these time averaged
(over years) profile shapes.  This is in agreement with the findings
by \citet{Dib08_glitches}, who compared the profile shapes for seven
(glitch-free) intervals of the total set~A plus~B.  We should note
that the PCA gain changed over time, what can offer an explanation for
some indications for small changes with time in the profiles as
discussed in \citet{Dib08_glitches}.

In order to exploit the maximal statistics, we created pulse profiles
using all available data collected with the PCA and HEXTE over nine
years (see Table~\ref{tab:rxsrxteobs} and
Sect.~\ref{sec:rxsrxte}). Before summing the profiles obtained at
different epochs we corrected for the above-mentioned PCA gain drift
in the conversion from channels to energy.  In Fig.~\ref{fig:rxsrxte8}
the \pca\ (panels A--E) and HEXTE (panels F--H) pulse profiles are
presented. The `lowest-significance' profiles above $\sim$15~keV
(Fig.~\ref{fig:rxsrxte8}\,D--H) have significances of $ 18.5\sigma,
4.8\sigma, 12.0\sigma, 6.4\sigma\, {\rm and}\, 4.3\sigma$,
respectively.

\subsubsection{Pulse profile changes with energy}
\label{sec:rxsppvar}

Comparing the profile shapes derived with INTEGRAL
(Fig.~\ref{fig:rxs8panel}) with those obtained with the PCA and HEXTE
in similar energy bands (Fig.~\ref{fig:rxsrxte8}), shows very good
agreement, \igr\ achieving the best statistics above 70 keV. Above
$\sim$ 20 keV there does not seem to occur significant changes in
morphology with energy.  The pulse profiles are statistically similar
which was tested using a combination of a Pearson $\chi^2$ \citep[see
  e.g. Sect.~11.2 of][]{Eadie71} and a Run test \citep[see
  e.g. Sect.~11.3.1 of][]{Eadie71}.

Below 20 keV, the situation is different: The \xmm\ and \pca\ profiles
show strong morphology changes as a function of energy. This has been
noted in a number of earlier publications
\citep[e.g.][]{Sugizaki97_1708,Israel01_1708,Kuiper06_axps}. In fact,
the morphology is changing faster with energy than can be seen in the
broad energy bins selected for Fig.~\ref{fig:rxs8panel} and
Fig.~\ref{fig:rxsrxte8}. In addition, there are systematic effects
which should be realized when comparing \xmm\ and PCA profiles.  For
example, in the measured energy band 2.2--3.9 keV
(Fig.~\ref{fig:rxsrxte8}\,A) one can see in the PCA profile a small
secondary peak at phase $\sim$0.8, which is absent in the
corresponding \xmm\ profile (Fig.~\ref{fig:rxs8panel}\,B).  This can
be explained with the significantly coarser energy resolution of the
PCA, accepting more events with photon energy above 4 keV in the band
with measured energies below 4 keV. For the same reason, this
secondary peak at phase 0.8 in the PCA 3.9--8.2 profile
(Fig.~\ref{fig:rxsrxte8}\,B) is higher than found in the corresponding
\xmm\ profile (Fig.~\ref{fig:rxs8panel}\,C), because this pulse
component is maximal for energies above 8 keV. Therefore, a more
accurate comparison can be made between the pulse characteristics
measured with \xmm\ and the \pca\ by performing phase-resolved
spectroscopy in which the energy resolution is automatically taken
into account (addressed in Sect.~\ref{sec:rxsphaseres}).

Nevertheless, the \xmm\ and
\pca\ profiles consistently show that the soft X-ray pulse has two
components which peak at phases $\sim$0.4 and $\sim$0.8. The pulse
peaking at phase $\sim$0.4 is spectrally softer than the pulse peaking
at phase $\sim$0.8. As mentioned above, the latter is not visible in
Fig.~\ref{fig:rxs8panel}\, A and B. This peak starts contributing
above $\sim$4~keV (Fig.~\ref{fig:rxs8panel}\,C and
Fig.~\ref{fig:rxsrxte8}\,B) and is the dominating pulse component
already above $\sim$8~keV up to and including the hard X-ray band.

The soft pulse (peak at phase $\sim$0.4) is gradually decreasing in
strength as a function of energy and seems to disappear entirely at
hard X-rays.  Finally, there is an indication for a third pulse
component, visible in the XMM profile below 2~keV where the \pca\ has
no sensitivity.  This component at phase $\sim$1.0 fills up the soft
pulse to a saw-tooth-like profile (Fig.~\ref{fig:rxs8panel}\,A), at
least, there is a sudden, significant change in pulse shape in
comparison to the profile above 2~keV.

\subsection{Energy spectra of the pulsed emission}
\label{sec:rxspulsedsp}

For \igr-ISGRI, \xmm-PN, \pca\ and RXTE-HEXTE pulsed spectra were
created by analyzing the pulse profiles following the procedure
described in Sect.~\ref{sec:rxsobs}. In Fig.~\ref{fig:rxstotpuls} an
example is given of how the pulsed counts are extracted as excess
counts above a flat background level. In this case for \rxs1708, the
$\sim$40\% pulsed emission in the total INTEGRAL energy band is
separated from the $\sim$60\% DC component. For \igr\ pulse profiles
for seven exponentially binned energy intervals within 20--300~keV
were created. For \xmm, \pca\ and HEXTE pulse profiles in 20
(0.5--12~keV), 15 (2.7-32.1~keV) and 5 (14.8--267.5~keV) energy
intervals were produced, respectively. In this section we first show
the total-pulsed phase-averaged spectra. Then we show for all three
missions the results of phase-resolved spectroscopy applied for three
broad phase bins selected in the \igr\ profile in
Fig.~\ref{fig:rxstotpuls}. Finally, we show phase-resolved pulsed
spectra for narrow ($\Delta \phi = 0.1$) phase bins for \pca\ and
\igr\ leading to the identification of three spectrally different
components which contribute to the total pulsed emission.

\subsubsection{Total-pulsed spectra}
\label{sec:rxsphaseav}

In Fig.~\ref{fig:rxstothigh} the time-averaged total-pulsed spectra
are shown for all four instruments using all available exposures. The
flux values from \xmm, \pca, RXTE-HEXTE and \igr\ are very consistent,
yielding a continuous broad-band spectrum from 0.5~keV up to almost
300~keV.  The \pca\ total-pulsed spectrum smoothly bridges the energy
gap between \igr\ and \xmm.  The RXTE-HEXTE spectrum is in agreement
with the \igr\ spectrum and also connects to the \pca\ spectrum.

For \pca\ we first created spectra for sets A and B separately, to
verify whether the time-averaged \rxs1708 spectra differ in the years
before and during the INTEGRAL observations. Within the statistical
errors the spectra appeared to be identical, in flux and shape. This
is in agreement with the recent findings of \citet{Dib08_glitches}, who
also concluded that the pulsed emission is very stable over the whole
time span from 1998 until 2006. Therefore, the combined spectrum with
maximal statistics using both sets of data is shown in
Fig.~\ref{fig:rxstothigh} and is further described below.

Fitting \igr, \pca\ and HEXTE simultaneously with two power-law
components yields an excellent fit for energies above 2.8 keV (see
Table~\ref{tab:rxstp}). The soft power law has a photon index
$\Gamma = 2.79 \pm 0.07$ and the hard power law has a photon index
$\Gamma = 0.86 \pm 0.16$ (\chir = 0.51, dof = 22). Fitting the
\igr\ spectrum separately it can be described by a single power-law
model with a photon index $\Gamma = 0.98 \pm 0.31$ (\chir = 0.70, dof =
5), which is nicely within errors of the combined \igr-RXTE fit.

\begin{table}[t]
\centering
\renewcommand{\tabcolsep}{1.7mm}
\caption[]{Spectral-fit parameters of the total-pulsed \igr, \xmm,
  \pca\ and RXTE-HEXTE fits (see Sect.~\ref{sec:rxsphaseav}).  The
  $N_{\rm H}$ is fixed to $1.47 \times 10^{22}$ cm$^{-2}$. The
  integrated fluxes are in units \ecs. $F_{0}$ is the normalization
  at 1 keV in units \pcsk. The subscripts s and h stand for `soft' and
  `hard' in the combined fit.
}

\begin{tabular}{lcc}

\vspace{-3mm}\\

\hline
\hline

\vspace{-3mm}\\
\multicolumn{2}{c}{\underline{\igr\ 20--270~keV}}\\
\vspace{-3mm}\\

$\Gamma_{\rm{\igr}}$ & $0.98 \pm 0.31$\\
$F_{20-150\,\rm{keV}}$ & $(2.53 \pm 0.40) \times  10^{-11}$\\
\vspace{-2mm}\\
\chir\, (dof)       & 0.70 (5)\\
$F_{20-270\,\rm{keV}}$ & $(4.89 \pm 1.12) \times  10^{-11}$\\

\vspace{-3mm}\\
\multicolumn{2}{c}{\underline{\xmm\ 0.5--2.8~keV$^\dagger$}}\\
\vspace{-3mm}\\

$\Gamma_{\rm{\xmm}}$ & $2.89 \pm 0.06$\\
$F_{0\,{\rm{\xmm}}}$ & $(1.95 \pm 0.07) \times  10^{-2}$\\
\vspace{-2mm}\\
\chir\, (dof)       & 0.46 (7)\\
$F_{0.5-2\,\rm{keV}}^\dagger$  & $(4.69 \pm 0.24) \times  10^{-11}$\\

\vspace{-3mm}\\
\multicolumn{2}{c}{\underline{\xmm\ 2.8--12~keV$^\dagger$}}\\
\vspace{-3mm}\\

$\Gamma_{\rm{\xmm}}$ & $2.87 \pm 0.12$\\
$F_{0\,{\rm{\xmm}}}$ & $(1.43 \pm 0.25) \times  10^{-2}$\\
\vspace{-2mm}\\
\chir\, (dof)       & 0.39 (6)\\

$F_{2-10\,\rm{keV}}^\dagger$  & $(1.21 \pm 0.05) \times  10^{-11}$\\



\vspace{-3mm}\\
\multicolumn{2}{c}{\underline{\pca+HEXTE+\igr\ 2.8--270~keV}}\\
\vspace{-3mm}\\

$\Gamma_{\rm{s}}$ & $2.79 \pm 0.07$\\
$F_{0\,{\rm{s}}}$ & $(1.34 \pm 0.11) \times  10^{-2}$\\
$\Gamma_{\rm{h}}$ & $0.86 \pm 0.16$\\
$F_{{\rm{h}}, 20-150\,\rm{keV}}$ & $(2.40 \pm 0.32) \times  10^{-11}$\\
\vspace{-2mm}\\
\chir\, (dof)             & 0.51 (22)\\
$F_{2-10\,\rm{keV}}$        & $(1.243 \pm 0.011) \times  10^{-11}$\\
$F_{10-20\,\rm{keV}}$       & $(0.333 \pm 0.010) \times  10^{-11}$\\
$F_{20-150\,\rm{keV}}$      & $(2.60 \pm 0.35) \times  10^{-11}$\\
$F_{20-270\,\rm{keV}}$      & $(5.16 \pm 1.06) \times  10^{-11}$\\

\hline

\end{tabular}
\flushleft$^\dagger$: Model includes fixed parameters $\Gamma_{\rm{h}}$ and
$F_{{\rm{h}}, 20-150\,\rm{keV}}$ of the combined RXTE-\igr\ spectrum
to correct for the contribution of the hard X-ray spectrum in the
\xmm\ band. The integrated flux is the model flux including the hard component.
\label{tab:rxstp}
\end{table}

To determine the model parameters of the 0.5--12.0~keV \xmm\ spectrum
the contribution of the hard power-law spectral component (significant
above $\sim$5~keV) were added with fixed parameters (as determined from
the \igr-RXTE fit) to the model and thereby effectively subtracted.
Fitting the 0.5--12.0~keV \xmm\ spectrum with a power-law yields a
rather poor fit with photon index $\Gamma = 3.08 \pm 0.03$ (\chir =
1.62, dof = 15).  This photon index is slightly softer than found for
\pca. However, the \xmm\ spectrum in Fig.~\ref{fig:rxstothigh} shows a
discontinuity between 2 and 3 keV, just below the \pca\ energy band.
Phase-resolved spectroscopy in the next section will show that this
discontinuity in the total-pulsed spectrum between 2 and 3 keV is
genuine.  Indeed, for energies above 2.8~keV an excellent fit can be
obtained with a power law with a photon index $\Gamma = 2.87 \pm
0.12$, which is in agreement with the \pca\ spectrum above 2.8~keV (in
flux and index, see Table~\ref{tab:rxstp}). The normalization of the
\xmm\ spectrum below 2.8~keV is higher than the one above 2.8~keV, but
the spectral shape is the same with an photon index $\Gamma = 2.89 \pm
0.06$.

The \xmm\ results imply an average pulsed fraction (defined as the pulsed
emission divided by the total emission) of (35.6$\pm$1.4)\% for the
2--10~keV band.  The \igr-RXTE fit gives a 20--150~keV pulsed flux of
$(2.56 \pm 0.40) \times 10^{-11}$ \ecs, resulting in an average pulsed
fraction of (39$\pm$6)\%. The broad-band 2--10~keV and 20--150~keV
pulsed fractions suggest a fairly constant pulsed fraction over the
total high-energy window. However, in Fig.~\ref{fig:rxstothigh} it can
be seen that the measured pulsed flux (204--300~keV) is on the
extrapolation of the power-law fit to the total spectrum. 
At these high energies the pulsed fraction could be as high as 100\%.

\subsubsection{Phase-resolved pulsed spectra}
\label{sec:rxsphaseres}

The morphology of the 20--270~keV \igr\ pulse profile
(Fig.~\ref{fig:rxstotpuls}) was used to define three broad phase
intervals: Ph\,I contains the DC level (green coloured in figures);
Ph\,II contains the shoulder in the \igr\ band (blue coloured in
figures); and Ph\,III contains the main pulse in the \igr\ band
(magenta coloured in figures). The phase interval boundaries are
listed in Table~\ref{tab:rxsph} and are indicated with vertical lines
and colour bars in Figs.~\ref{fig:rxstotpuls}, \ref{fig:rxs8panel} and
\ref{fig:rxsrxte8}.  In the latter figures is also visible how these
phase intervals relate to the profile shapes at energies below 10 keV.

\begin{table}[!thb]
\centering
\renewcommand{\tabcolsep}{1.7mm}
\caption[]{Selected phase intervals for extraction of high-energy
  spectra, using the pulse-shape morphology of the \igr-IBIS-ISGRI
  20--270~keV pulse profile (Fig.~\ref{fig:rxstotpuls}).}

\begin{tabular}{lll}

\vspace{-3mm}\\

\hline
\hline
\vspace{-3mm}\\
 & Phase interval & Component \\

\hline
\vspace{-3mm}\\
Ph\,I   & $\lbrack0.95, 0.25\rangle$ & DC level \igr\ profile\\
Ph\,II  & $\lbrack0.25, 0.65\rangle$ & Shoulder in \igr\ band \\
Ph\,III & $\lbrack0.65, 0.95\rangle$ & Main hard \igr\ pulse\\

\hline
\end{tabular}

\label{tab:rxsph}
\end{table}

\begin{table}
\centering
\renewcommand{\tabcolsep}{1.7mm}
\caption[]{Spectral-fit parameters for Ph\,III \igr, \xmm, \pca\ and
  RXTE-HEXTE fits (see Sect.~\ref{sec:rxsphaseres}).  The $N_{\rm H}$
  is fixed to $1.47 \times 10^{22}$ cm$^{-2}$. $kT$ and $E_{{\rm
      break}}$ are in units keV. $F_{0}$ is the normalization at 1
  keV in units \pcsk. The integrated fluxed are in units \ecs.
}

\begin{tabular}{lcc}

\vspace{-3mm}\\

\hline
\hline
\vspace{-3mm}\\
\multicolumn{2}{c}{\underline{\xmm+\pca+HEXTE+}}\\
\multicolumn{2}{c}{\underline{\igr\ 0.5--270~keV}}\\
\vspace{-3mm}\\

$kT$                   & $0.23 \pm 0.02$\\
$F_{0\,kT}$             & $(2.82 \pm 0.64) \times  10^{-5}$\\
$\Gamma_{\rm{I}}$       & ${\bf -}0.55 \pm 0.14$\\
$E_{\rm{break}}$        &  $4.92 \pm 0.11$\\
$\Gamma_{\rm{II}}$      & $1.77 \pm 0.09$\\
$F_{0\,}$               & $(2.05 \pm 0.52) \times  10^{-5}$\\
$\Gamma_{\rm{III}}$     & $0.02 \pm 0.50$\\
$F_{20-150\,\rm{keV}}$   &$(0.86 \pm 0.22) \times 10^{-11}$ \\
\vspace{-2mm}\\
\chir\, (dof)          & 0.90 (29)\\
$F_{0.5-2\,\rm{keV}}$ (\ecs) & $(0.19 \pm 0.07) \times  10^{-11}$\\
$F_{2-10\,\rm{keV}}$ (\ecs)  & $(0.23 \pm 0.07) \times  10^{-11}$\\
$F_{20-150\,\rm{keV}}$(\ecs) &$(1.54 \pm 0.18) \times 10^{-11}$ \\
$F_{20-270\,\rm{keV}}$(\ecs) &$(3.74 \pm 0.65) \times  10^{-11}$ \\

\vspace{-3mm}\\
\multicolumn{2}{c}{\underline{\igr\ 20--270~keV}}\\
\vspace{-3mm}\\

$\Gamma_{\rm{\igr}}$ & $1.09 \pm 0.16$\\
$F_{20-150\,\rm{keV}}$(\ecs) &$(1.53 \pm 0.12) \times 10^{-11}$ \\
\vspace{-2mm}\\
\chir\, (dof)       &  1.15 (5)\\
$F_{20-270\,\rm{keV}}$(\ecs) &$(2.81 \pm 0.42) \times  10^{-11}$ \\

\hline

\end{tabular}

\label{tab:rxsph3}
\end{table}

\begin{figure*}
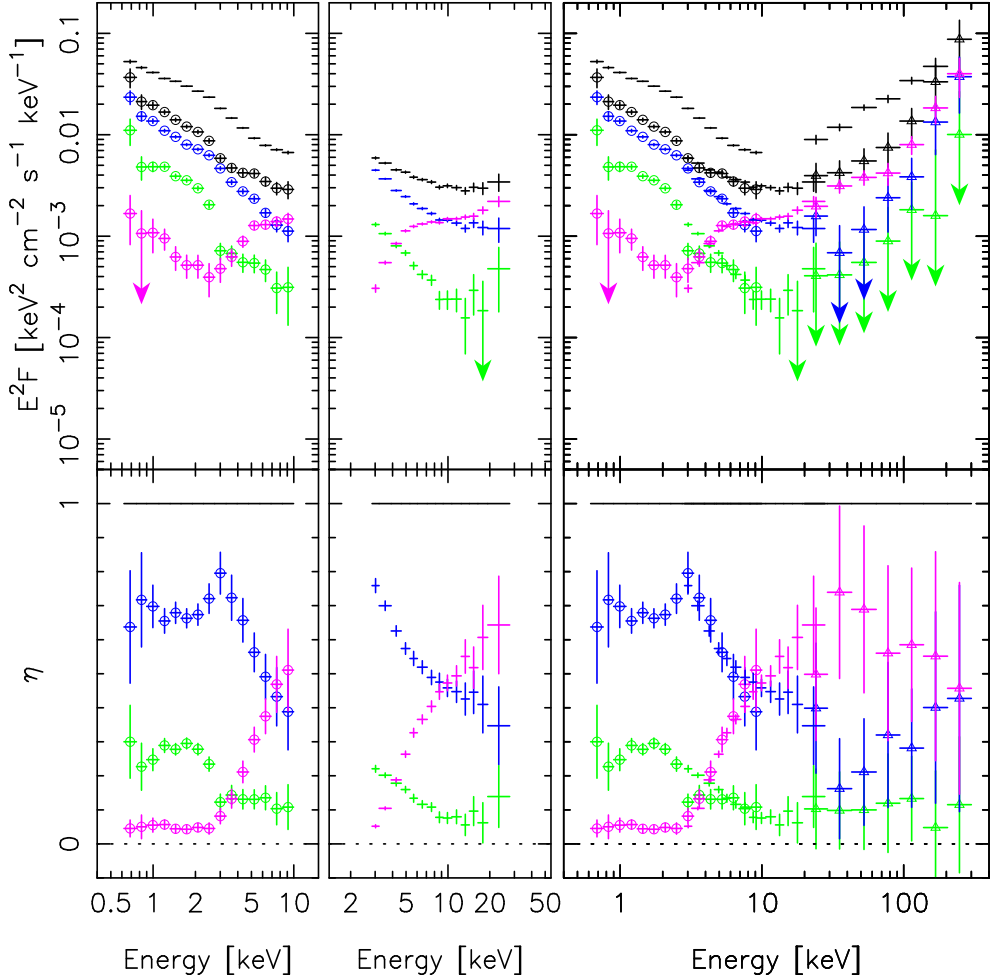

\centering
\begin{minipage}[c]{0.3\textwidth}
\psfig{figure=09772fig08a.ps,height=13cm,angle=0,clip=t}
\end{minipage}
\begin{minipage}[c]{0.3\textwidth}
\psfig{figure=09772fig08b.ps,height=13cm,angle=0,clip=t}
\end{minipage}%
\begin{minipage}[c]{0.3\textwidth}
\psfig{figure=09772fig08c.ps,height=13cm,angle=0,clip=t}
\end{minipage}
\caption{Phase-resolved results for \xmm\ and \pca\ are presented in
  the left and center panels, respectively. These data and the
  \igr\ (overlapping) results are shown together in the right
  panels. Total (except for RXTE), total-pulsed and phase-resolved
  pulsed spectra are presented in the top panels. The total spectra
  and total-pulsed spectra are plotted in black. The phase resolved
  spectra for Ph\,I, Ph\,II and Ph\,III (see Table~\ref{tab:rxsph})
  are plotted in green, blue and magenta, respectively. The
  \xmm\ pulsed spectra are also indicated with open circle symbols
  while the \igr\ pulsed spectra are indicated with a triangle
  symbols. The data points with arrows indicate that the flux values
  have significances less than 1.5$\sigma$ and the positive 1$\sigma$
  error is drawn. In the bottom panels $\eta$ -- defined as the fraction
  of the total pulsed emission within a phase interval -- is shown in
  the same colour scheme as the spectra. }
 \label{fig:rxseta}
\end{figure*}
%

In the upper panels of Fig.~\ref{fig:rxseta} the resulting
pulsed-emission spectra for these three phase intervals are shown, in
each case together with the total-pulsed spectrum, and for \xmm\ and
\igr\ also with the total-emission spectra; \xmm\ in the upper-left
panel; \pca\ in the top-center panel; and in the top-right panel \xmm,
\pca\ and \igr\ together.  In the \pca\ analysis we have again
verified that the three phase-resolved spectra for Set~A and Set~B are
fully consistent: there is no sign for long-term variability in shape
and flux down to the 5\%-level.

We first focus on the pulsed-emission spectrum of Ph\,III which
contains the main hard \igr\ pulse. This Ph\,III spectrum follows the
total-pulsed spectrum in the \igr\ band, as expected. However, Ph\,III
shows below 10~keV a completely different spectral behaviour with
respect to the total-pulsed spectrum. Below 10~keV where the
total-pulsed spectrum turns up towards lower energies, this spectrum
keeps going down steeply.
The pulsed emission of Ph\,III reaches a minimum in luminosity around
$\sim$2.5~keV. At lower energies it turns up again. This remarkably
spectral shape is measured consistently with \xmm\ and \pca, as can be
seen in the right panel of Fig.~\ref{fig:rxseta} (magenta coloured
data points).  Therefore, we can exclude a systematic problem as cause
of this appearance. It is also remarkable how smoothly the combined
\xmm\ and nine-year-average \pca\ spectra join the four-year-average
\igr\ spectrum.  This is strong evidence for a very stable
configuration of the production sites of the high-energy emission in
the magnetar magnetosphere over almost a decade.

It is obviously not straightforward to fit the spectrum of Ph\,III. We
attempted to make an empirical description to `quantify' the
discontinuities using all data points from \igr, \pca\ and \xmm. We
succeeded in describing the spectrum over the whole 0.5--270~keV band
with a black body and three power-law segments (i.e. one broken power
law plus a power law; \chir = 0.90 dof = 29, see
Table~\ref{tab:rxsph3}). The total fit contains a soft thermal
black-body component below 2~keV ($kT = 0.23 \pm 0.02$ keV). This
component can be due to a tail of the soft pulse which peaks in
Ph\,II. Above $\sim$2~keV the spectrum drastically hardens. The
spectrum up to the break energy $E_{{\rm break}}= 4.92 \pm 0.11$ keV
is extremely hard with a photon index $\Gamma = {\bf-}0.55 \pm
0.14$. After the break a softer power law is required with photon
index $\Gamma = 1.77 \pm 0.09$. Finally, in order to account for the
harder \igr\ spectrum a third power law is required with photon index
$\Gamma = 0.02 \pm 0.50$, which is dominating above $\sim$60~keV. It
is obvious that very different production processes contribute in this
phase interval (Ph\,III) to the high-energy (pulsed) spectrum.

Fitting only the \igr\ spectrum with a power law, a photon index
$\Gamma = 1.09 \pm 0.16$ gives the best description.
(Table~\ref{tab:rxsph3}).

Also the spectral shape of the pulsed emission in Ph\,II (centered on
the soft pulse peaking at phase $\sim$0.4, blue coloured in figures)
is very different from that of the total-pulsed spectrum. 

The independent \xmm\ and \pca\ spectra are again fully consistent. A
broad soft component below about 8 keV turns in the RXTE band above 8
keV into a component with power-law index $\sim$2, but is not detected
around 20 keV. The flux measurements close to energies as high as 100
keV are likely not related to the softer X-ray components and can be
explained as emission from a tail of the main hard-X-ray \igr\ pulse
in Ph\,III. No sensible model fits can be made to the total combined
Ph\,II spectrum. A good fit can be obtained considering only
\xmm\ data, namely with a broken power-law with break energy $E_{{\rm
    break}}= 2.49 \pm 0.15$ keV, consistent with the energy where in
the total pulsed spectrum the discontinuity was found. Below the break
energy the power-law model has a photon index $\Gamma = 2.83 \pm
0.04$, above this energy $\Gamma = 3.38 \pm 0.06$.

Ph\,I contains the left wing of the soft pulse below 8 keV, and
harbours the DC level in the \igr\ band (green coloured in
figures). Therefore, no significant flux is measured at hard
X-rays. Also in this case no sensible model combination (black body,
logparabola, power law) can describe the combined
\xmm--\pca\ spectrum. Furthermore, the \xmm\ spectrum exhibits again a
drastic drop in flux between 2 and 3 keV. The fact that the \pca\ data
points between 2 and 3 keV do not follow this sudden drop, might be
explained with the coarser energy resolution of \pca, combined with a
strong gradient in the sensitive area and underlying photon spectrum.

\begin{figure}[t]
\psfig{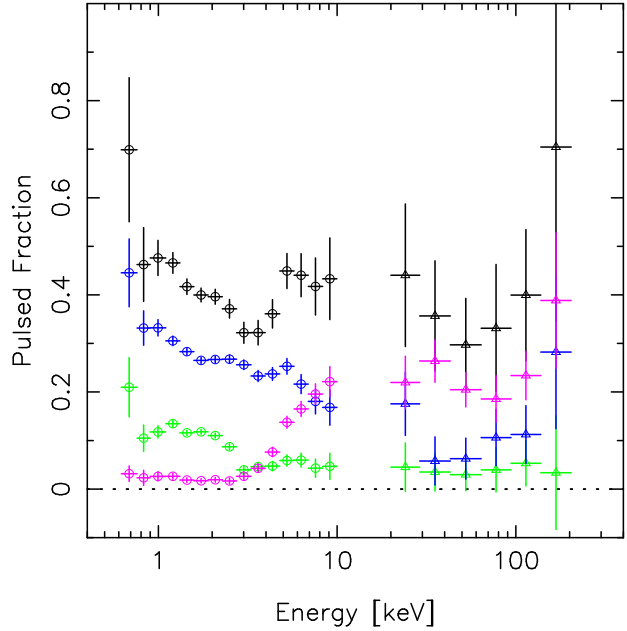}
\caption{The pulsed fractions of \rxs1708\ as a function of energy
  using \xmm\ below 10~keV and \igr\ above 20~keV. The total pulsed
  fraction is drawn in black and plotted in colour are the pulsed
  fractions for the three phase intervals (see Table~\ref{tab:rxsph}).
\label{fig:rxspf}}
\end{figure}

A different way to study the relative contributions of the pulsed
components in the three selected phase intervals as a function of
energy is by plotting the $\eta$ parameter, defined as the fraction of
the total-pulsed emission. It shows more clearly discontinuities in
the dependence on energy. In the bottom panels of
Fig.~\ref{fig:rxseta} $\eta$ is presented for the different
observations. The bottom panel for \xmm\ shows that the Ph\,III
contribution to the pulsed emission is a few percent up to
$\sim$2.5~keV. Above 2.5~keV the contribution increases as a function
of energy to $\sim$50\% at 10~keV. This picture is fully confirmed by
RXTE, which shows that $\eta$ is still increasing above 10~keV up to
$\sim$60\% around 30~keV and stabilizes around this value in the
\igr\ band.

The opposite can be seen for the contribution to the pulsed flux from
Ph\,II. A high $\eta$ of $\sim$70\%, fairly constant up to $\sim$3~keV
from where the contribution abruptly decreases to $\sim$40\% at
10~keV. Also this decline is observed with RXTE. The behaviour in the
\igr\ band is not clear.  Finally, Ph\,I shows a somewhat similar
trend as Ph\,II with a significant contribution below $\sim$3~keV of
$\sim$28\%, which decreases to $\sim$8\% above $\sim$8~keV and only
upper limits in the \igr\ band.

For \xmm\ and \igr\ the pulsed fraction is shown in
Fig.~\ref{fig:rxspf}. The pulsed fraction (total-pulsed emission
divided by total emission) of \rxs1708\ is around $\sim$40\% from soft
to hard X-rays (see also Sect.~\ref{sec:rxsphaseav}). However, we can
note here, that the pulsed fraction is not constant with energy, but
shows significant variations with energy below 10 keV. In the
\igr\ band the error bars are too large to draw firm conclusions. In
Fig.~\ref{fig:rxspf} the pulsed fractions for the three phase
intervals are also shown (i.e. emission in Ph\,I, II or III divided by
total emission).  As was also clear from the dependence of $\eta$ on
energy in Fig.~\ref{fig:rxseta}, the contribution of the hard X-ray
pulse increases with energy above $\sim$3~keV, while the pulsed
fractions of the other two phase intervals are decreasing with energy.

\subsubsection{Narrow-band phase resolved pulsed spectra from RXTE}
\label{sec:rxsrxtephase}
\begin{figure*}[!ht]
\centering
\begin{minipage}{1.8\columnwidth}
  \psfig{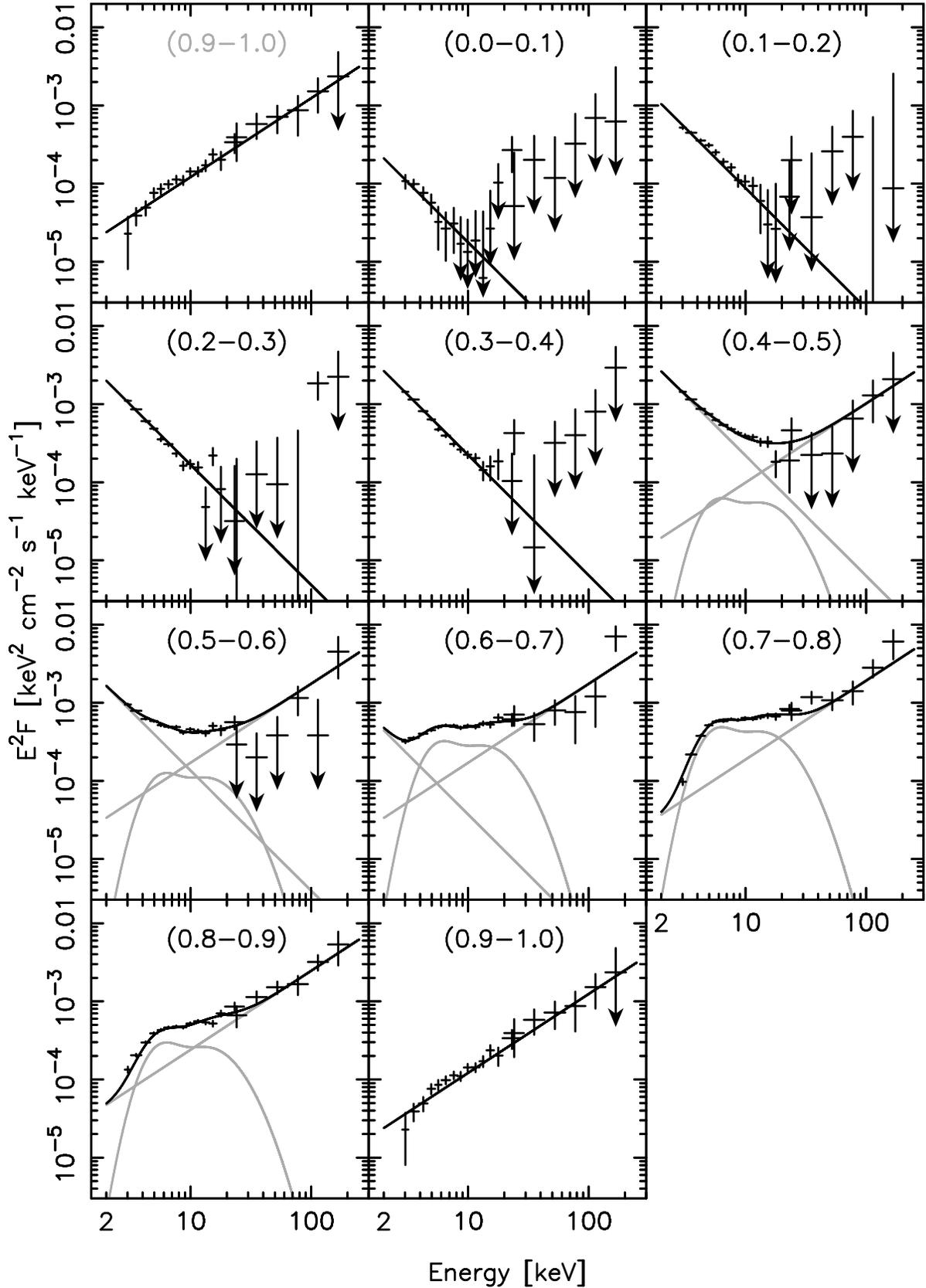}
\end{minipage}
\caption{Narrow-band phase-resolved spectroscopy of \rxs1708 using
  \pca\ and \igr-ISGRI data. Each panel shows the spectrum of a phase
  interval with $\Delta\phi = 0.1$. The phase intervals are indicated
  in the panels. The data points with arrows indicate that the flux
  values have significances less than 1.5$\sigma$. The positive
  1$\sigma$ errors are drawn. Total spectral fits and the three
  components (when contributing significantly) are drawn in black and
  grey, respectively (See Sect.\ref{sec:rxsrxtephase} and
  Fig.~\ref{fig:rxs10spec}).
\label{fig:rxs10spec}}
\end{figure*}
\begin{figure}
\psfig{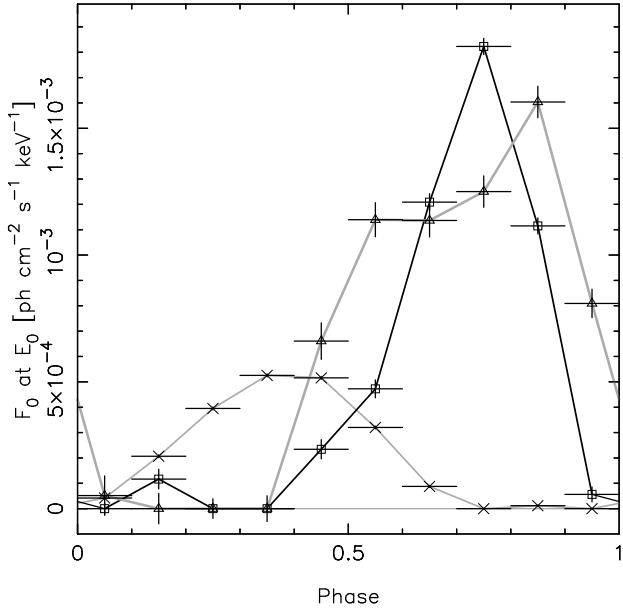}
\caption{Phase distributions of the normalizations $F_0$ of the three
  spectral-model components (see Table~\ref{tab:rxsmodpar}) taken at
  pivot energy $E_0 = 15.0433$ keV for the 10 narrow phase-interval
  spectra. Indicated with crosses and a thin grey line are the
  normalizations of the soft power-law component, with triangles and
  thick grey line the hard power-law component and with squares and a
  black line the composite spectral component.
\label{fig:rxsmodpar}}
\end{figure}

The phase-resolved spectroscopy presented in
Sect.~\ref{sec:rxsphaseres}, selecting three broad phase intervals,
unambiguously revealed that spectrally different components contribute
to the total pulsed emission of \rxs1708. A similar conclusion was
earlier reached for AXP~\0142 \citep{denHartog08_0142}.  However, the
situation appears to be much more complex for \rxs1708.  The
\xmm\ spectra below $\sim$2 keV suggest a soft thermal component,
which is not unexpected. However, this component can not be easily
separated from components with apparently complex spectral shapes
contributing to the total pulsed emission above 2 keV.  Therefore, we
exploited the high statistics of the total $\sim$600~ks of \pca\ data
to explore in more detail the spectral characteristics as a function
of phase above 2 keV and the connection to the \igr\ band. We repeated
the phase-resolved spectroscopy but now for 10 narrow bins each
covering 0.1 in phase.  Applying the method described in
Sect.~\ref{sec:rxsrxte} the 10 phase-resolved spectra have been
extracted and are shown in Fig.~\ref{fig:rxs10spec}.

The variations in spectral shape with phase are drastic. In the first
4 phase bins between phase 0.0 and 0.4, we see a spectral tail of the
soft component peaking in energy in the \xmm\ band below 1 keV. These
spectra can be accurately fitted for energies above 2.8 keV with a
single power-law model. The index for the best fit to the sum of the
spectra between phase 0.0 and 0.3 is $\Gamma = 3.54 \pm 0.01$.  In the
phase bins 0.4--0.5 and 0.5--0.6 a significant hardening of the
spectra sets in for energies between about 5 keV and 20 keV. In the
following phase bins 0.6--0.7 and 0.7--0.8, getting closer to the
phase of the main INTEGRAL pulse, evidence is visible for the presence
of the hard power-law component with index $\sim$1.0 above 20
keV. This latter hard power-law spectrum is dominating the energy
budget for the last three phase bins. More surprisingly, in the two
phase windows between 0.7 and 0.9, centered on the INTEGRAL main
pulse, a very hard part of the spectrum is apparent in a narrow energy
interval above $\sim$2 keV.  This is most extreme for the phase
interval 0.7--0.8 with a `local' photon index $\Gamma = -1.49 \pm
0.21$ (notice the sign!) up to a break energy $E_{{\rm break}} = 4.69
\pm 0.04$.  Finally, the spectrum for phase interval 0.9--1.0, the
last fraction of the trailing wing of the INTEGRAL main pulse, can be
fitted with a single power law with best-fit photon index $\Gamma =
0.99 \pm 0.05$. Moving only 10\% in phase to the next phase interval
0.0--0.1, the hard spectrum `switches' and becomes again very soft
with $\Gamma = 3.58 \pm 0.34$.

\begin{table}[!t]
\centering
\renewcommand{\tabcolsep}{1.7mm}
\caption[]{Model parameters of the three spectral components that are
  used to fit the 10 narrow-band phase-resolved \pca\ and \igr-ISGRI
  spectra. The three components are a soft and a hard power-law
  component, and a curved component which is the sum of two
  logparabolic functions. For each component $F_{0}$ (in units \pcsk)
  is the normalization taken at the pivot energy $E_0 = 15.0433$ keV
  which is plotted as a function of phase in
  Fig.~\ref{fig:rxsmodpar}.}

\begin{tabular}{c}

\vspace{-3mm}\\

\hline
\hline
\vspace{-3mm}\\

\underline{Power-law components $F = F_0 \times
     \left(\frac{E}{E_0}\right)^{-\Gamma}$}\\
Soft power law  $\Gamma_s = 3.54$\\
Hard power law  $\Gamma_h = 0.99$\\
\vspace{-3mm}\\

\underline{Composite component}\\
\vspace{-3mm}\\
\underline{ $F = F_{0} \times
    \left[k \cdot \left(\frac{E}{E_{0}}\right)^{-\alpha_1 -\beta_1
        \cdot{\rm ln}\left(\frac{E}{E_0}\right)} +
      \left(\frac{E}{E_{0}}\right)^{-\alpha_2 -\beta_2 \cdot{\rm
          ln}\left(\frac{E}{E_0}\right)}\right]$ }\\
\vspace{-3mm}\\
$k = 7.2632 \times 10^{-3}$   \\
$\alpha_1 = 11.635866$\\
$\beta_1 = 4.867669$\\
$\alpha_2 = 2.3099390$\\
$\beta_2 = 1.6262881$\\

\vspace{-3mm}\\
\hline
\vspace{-3mm}\\

\hline

\end{tabular}
\label{tab:rxsmodpar}
\end{table}

From Fig.~\ref{fig:rxs10spec} it is evident that different spectral
components contribute to the total pulsed emission and dominate in
different phase intervals. Two components can clearly be identified: A
soft power-law component with index $\Gamma \sim 3.6$ and a hard
power-law component with index $\Gamma \sim 1.0$. Then there appears
to be a third component, dominating in the energy band $\sim$3 to 20
keV for phases between 0.7 and 0.9. A clear characteristic of this
spectral component is a very steep rise above 2.3 keV ($\Gamma \sim
-1.5$).  We investigated what shape such a third component should
have in order to explain, together with two power-law components, all
spectra in Fig.~\ref{fig:rxs10spec}. This shape should give a clear
constraint for any proposed production mechanism. 

We followed the following procedure. The hard power-law component was
adopted from the best fit to the spectrum for phase 0.9--1.0: $\Gamma
= 0.99$. Then we investigated empirically what spectral shape,
together with the power-law model with $\Gamma = 0.99$, can accurately
describe the spectrum in phase interval 0.7--0.8. This turned out to
be the sum of two logparabolic functions of which the parameters are
given in Table~\ref{tab:rxsmodpar}.  Finally, for the index of the
soft power-law model we adopted the best fit to the summed spectrum in
the phase interval 0.0--0.3: $\Gamma = 3.54$.

We now attempted to describe the measured spectral distributions in
the 10 narrow phase intervals in terms of just these three model
shapes with free normalizations. Interestingly, the resulting
total fits are very satisfactory for all narrow phase intervals (see
Fig.~\ref{fig:rxs10spec}).  The normalizations of the three models are
shown in Fig.~\ref{fig:rxsmodpar}. Effectively, this figure shows the
pulse profile of each spectral component contributing to the
total-pulsed spectrum, i.e. a complete decoupling based on the
measured spectral and timing characteristics. In
Fig.~\ref{fig:rxs10spec} we show for all 10 spectra the contributions
of the three spectral components to the model fit, but only when
components were statistically required to reproduce the measured
spectra. Fig.~\ref{fig:rxsmodpar} together with
Fig.~\ref{fig:rxs10spec} show now the characteristics of the three
components which can reproduce together the total pulsed emission, the
pulse shapes and their spectra. Furthermore, we have shown above that
our results indicate that these characteristics are invariant over
almost a decade of observations.

\section{Summary}
\label{sec:rxssum}

In this paper we have presented new and more detailed characteristics
of \rxs1708 in the hard X-ray/soft gamma-ray regime ($>$20 keV) using
\igr-ISGRI and SPI data.  Following the approach in our earlier paper
on AXP \0142 \citep{denHartog08_0142} we extended the energy window to
lower energies to obtain a broader high-energy view using archival
RXTE and \xmm\ observations. The emphasis of this work on broad-band
phase-resolved spectroscopy was aimed at the identification of
distinctly different components contributing to the total-pulsed
emission.  What follows is a summary of the main results.

\subsection{Total high-energy emission above 20 keV of \rxs1708}

1) Using all available \igr\ data taken with \igr\ ISGRI during
2003--2006 -- adding up to 5.2 Ms effective on-source exposure -- the
time-averaged 20--175~keV spectrum can be described by a power-law
with photon index $\Gamma=1.13 \pm 0.06$.  The luminosity is
$1.34 \times 10^{35}$ \es (20--175~keV) adopting a distance of 3.8 kpc
({\it Sect.~\ref{sec:rxsigrtot}, Fig.~\ref{fig:rxstothigh}, Table
  \ref{tab:rxsfits}}).

\noindent 2) There is no indication for a spectral break in the
\igr\ spectrum up to energies of $\sim$ 175 keV.  Including in the
spectral fit earlier published COMPTEL flux upper limits for energies
above 750 keV \citep{Kuiper06_axps} and assuming a logparabolic shape
for the spectrum above 20 keV, the maximum luminosity is found for an
energy of $\sim$290~keV ({\it Sect.~\ref{sec:rxsigrtot},
  Fig.~\ref{fig:rxstothigh}}).

\subsection{Long-term variability in the total emission of \rxs1708}

1) In the INTEGRAL data no significant long-term time variability for
energies above 20 keV is found on time scales of one year and half a
year in total flux and spectral index. There is only one measurement
with a $\sim$3$\sigma$ indication for a flux increase in nine
observations. The total flux (20--150\,keV) and the spectral index are
stable within 22\% and 18\% (both 1$\sigma$), respectivily ({\it
  Sect.~\ref{sec:rxsigrvar}, Figures~\ref{fig:rxs4spec},
  \ref{fig:rxscontours} \& \ref{fig:rxsfluxes}}).
  
\noindent 2) Our analysis of all available ISGRI data rejects the
claim by \citet{Gotz07_1708} of correlated variability between the
hard X-ray emission measured with ISGRI and soft X-ray emission from
\rxs1708 in relation to a (candidate) glitch ({\it
  Sect.~\ref{sec:rxsigrvar}, Fig.~\ref{fig:rxsfluxes}}).

\subsection{Pulse profiles}

1) ISGRI measures significant pulse profiles for energies up to
270~keV ({\it Sect.~\ref{sec:rxsigrxmmpp},
  Figures~\ref{fig:rxstotpuls} \& \ref{fig:rxs8panel}}).

\noindent 2) High-energy pulse profiles measured with \xmm, \pca\ and
HEXTE and \igr-ISGRI over the broad energy band 0.5 keV up to 270 keV
are consistent in morphology when measured in the same differential
energy bands, independent of the epoch of the observation; apparent
differences are either due to statistics or due to differences in
energy response ({\it Sect.~\ref{sec:rxspp},
  Figures~\ref{fig:rxs8panel} \& \ref{fig:rxsrxte8}}).

\noindent 3) Fast morphology changes with energy are visible in the
pulse profiles below $\sim$20~keV, \citep[earlier noted by
  e.g.][]{Sugizaki97_1708,Israel01_1708,Kuiper06_axps} In the
\igr\ band the pulse profiles do not show morphology changes ({\it
  Sect.~\ref{sec:rxspp}, Figures\ref{fig:rxs8panel} \&
  \ref{fig:rxsrxte8}}).

\noindent 4) Three different pulse components are recognized in the
pulse profiles: a) A hard pulse peaking around phase 0.8; which starts
contributing to the pulse profiles above $\sim$4~keV, b) a softer
pulse peaking around phase 0.4; which is not apparent at hard X-rays,
c) a very soft pulse component below 2~keV for which there is an
indication around phase 1 where it fills the pulse profile to a
saw-tooth-like shape ({\it Sect.~\ref{sec:rxsigrxmmpp},
  Fig.~\ref{fig:rxs8panel}}).

\subsection{Total-pulsed spectra}

1) ISGRI measures the phase-averaged total-pulsed spectrum up to
270~keV which can be described with a power law with photon index
$\Gamma = 0.98 \pm 0.31$ ({{\it Sect.~\ref{sec:rxsphaseav},
    Fig.~\ref{fig:rxstothigh}}).

\noindent2) The average pulsed fraction in the \igr\ band is
39\%. Note, that the pulsed flux measured above 200~keV is on the
extrapolation of the total spectrum, consistent with being for 100\%
pulsed. Below 10 keV, the pulsed fraction as measured with \xmm\ is
not constant but varies between $\sim$31\% and $\sim$48\% in a complex
way. ({\it Sect.~\ref{sec:rxsphaseav}, Figures~\ref{fig:rxstothigh} \&
  \ref{fig:rxspf} }).

\noindent3) The total-pulsed spectrum from 0.5~keV up to 270~keV is
very stable over time-scales of years. Firstly, the pulsed-emission
spectra (2.7--32.1 keV) of two sets of RXTE-PCA data taken six years
before and three years during the \igr\ operations, respectively,
appeared to be statistically consistent within a few percent
\citep[see also][]{Dib08_glitches}.  Secondly, the summed RXTE-PCA
total-pulsed spectrum joins smoothly that from \xmm\ at lower
energies, as well as those from RXTE-HEXTE and \igr\ at higher energies
({\it Sect.~\ref{sec:rxsphaseav}, Fig.~\ref{fig:rxstothigh}}).

\noindent4) Fitting \igr, \pca\ and HEXTE simultaneously with two
power-law components yields an excellent fit for energies above 2.8
keV with a soft power-law model with photon index $\Gamma = 2.79 \pm
0.07$ and a hard power-law model with photon index $\Gamma = 0.86 \pm
0.16$ (\chir = 0.51, dof = 22). ({\it Sect.~\ref{sec:rxsphaseav},
  Table~\ref{tab:rxstp}})

\noindent5) The \xmm\ pulsed spectrum above 2.8 keV is fully
consistent in flux and power-law index with the corresponding
\pca\ spectrum, however, there is a discontinuity in the
\xmm\ spectrum, a drop in flux, between 2 and 3~keV ({\it
  Sect.~\ref{sec:rxsphaseav}, Fig.~\ref{fig:rxstothigh}}).

\subsection{Phase-resolved pulsed spectra}

1) The time-averaged phase-resolved spectra, selecting three broad
phase intervals or narrow 0.1-phase-wide intervals, are connecting
very smoothly over the total 0.5--300 keV energy band from instrument
to instrument: Another confirmation of the long-term many-year
stability of the geometry and production mechanisms responsible for
the emission ({\it Sect.~\ref{sec:rxsphaseres}, Fig.~\ref{fig:rxseta}
  \& \ref{fig:rxs10spec}}).

\noindent 2) The \xmm, \pca\ and \igr\ phase-resolved spectra,
selecting three broad phase intervals, have vastly different and
complex spectral shapes, constituting together the total-pulsed
spectrum ({\it Sect.~\ref{sec:rxsphaseres}, Fig.~\ref{fig:rxseta}}).

\noindent 3) Ph\,III (which contains the main hard \igr\ pulse) shows
hardly any contribution to the pulsed spectrum for energies below
2.5~keV (i.e. low $\eta$). This part of the spectrum can be described
by a black-body with a temperature $kT$ of $0.23 \pm 0.02$\,keV. Above
2.5~keV, the energy where the Ph\,III pulsed emission reaches its
minimum luminosity, the spectral shape becomes very complex: The
spectrum first hardens dramatically (and $\eta$ increases) with a
photon index $\Gamma = -0.55 \pm 0.14$ up to a break energy of $\sim
5$\,keV, consistently measured with \xmm\ and \pca. After the break
the spectrum softens to a power-law with photon index $\Gamma = 1.77
\pm 0.09$. In the \igr\ band it hardens again to a photon index
$\Gamma = 0.86 \pm 0.22$ ({\it Sect.~\ref{sec:rxsphaseres},
  Fig.~\ref{fig:rxseta}, Table~\ref{tab:rxsph3}}).

\noindent 4) The spectra of Ph\,I, DC-level in \igr\ pulse profile,
and Ph\,II, `shoulder' in \igr\ profile and main soft-X-ray pulse,
also appear to be complex. Both spectra are soft with maximum
luminosity below 1 keV and exhibit discontinuities around 2.8~keV, the
energy where we noted a drop in flux in the total-pulsed spectrum of
\xmm. The emissions seem to vanish above 20 keV ({\it
  Sect.~\ref{sec:rxsphaseres}, Fig.~\ref{fig:rxseta}}).

\noindent 5) Narrow-band phase-resolved spectroscopy in 10 phase
bins reveals the cause of the complex behaviour of the emission
spectra in the broader phase bins. The 10 spectra show gradual and
sudden changes as a function of phase. The spectral complexity has
become even clearer and is more extreme than seen in the broad-band
phase-resolved spectra ({\it Sect.~\ref{sec:rxsrxtephase},
  Fig.~\ref{fig:rxs10spec}}).

\noindent 6) Each of the 10 pulsed spectra can be accurately described
with a sum (free normalizations) of three components with very
different spectral shapes: a) a soft power-law component with a photon
index $\Gamma = 3.54$, b) a hard power-law component with a photon
index $\Gamma = 0.99$, and c) a peculiar curved component which has
been approximated by the sum of two logparabolic functions and
contributes significantly between 4 and 20 keV ({\it
  Sect.~\ref{sec:rxsrxtephase}, Fig.~\ref{fig:rxs10spec},
  Table~\ref{tab:rxsmodpar}}).

\noindent 7) The phase distributions of the normalizations to the
three spectral components represent three decoupled pulse profiles
which together constitute the total pulse profile. The soft component
shows a single peak around phase 0.4. The hard component shows a broad
peak with a significant contribution from phase 0.4 to 1. The curved
component shows a narrower peak ($\sim$0.25 in phase FWHM) which has its
maximum in phase interval 0.7--0.8 ({\it Sect.~\ref{sec:rxsrxtephase},
  Fig.~\ref{fig:rxsmodpar}}).


\section{Discussion}
\label{sec:rxsdisc}

The detailed analysis performed in this work on the high-energy
emission (0.5--300 keV) from \rxs1708 led to similar conclusions as
were reached earlier for AXP \0142 \citep{denHartog08_0142}.  Namely,
that genuinely different pulse components with different spectra
contribute to the total-pulsed emission.  In the case of \rxs1708 not
only the variation with energy of the structure of the pulse profiles
could be used to identify these different components, similar to the
case of \0142, but also the apparent fine structure in the
phase-resolved spectra provided an efficient means to disentangle the
pulsed emission into different components.

In \citet{denHartog08_0142} we discussed the implications of the
derived high-energy characteristics of \0142 for models developed in
recent years for explaining the non-thermal, luminous hard X-ray
emission from AXPs: 1) a quantum electro-dynamic model (QED) by
\citet{Heyl05_qed,Heyl05_qed2}, 2) a corona model by \citet{BT07},
further elaborated by \citet{Lyubarsky07_corona} and 3) a resonant
upscattering model by \citet{Baring07_london}. We concluded that all
three models have difficulties in explaining many features of the
detailed results \citep[See the discussion in the paper
  by][]{denHartog08_0142}.  A confrontation of these models with our
new results for \rxs1708 leads to similar conclusions.  It is apparent
that any attempt to explain the high-energy emission (0.5--300 keV)
should consider a stable (at least over a decade) three-dimensional
geometry taking all relevant angles (the angle between the magnetic
and the spin axis and the viewing angle) into account, as well as the
physical production processes taking place on the surface of the
neutron star and at different sites in the atmosphere and
magnetosphere.

For \rxs1708 our results reveal a clear separation in phase, and
therefore also of the production sites, of all pulsed hard X-ray
emission above 20 keV with photon index $\Gamma \sim 1.0$ from the
pulsed soft thermal emission below 4 keV. This soft emission can be
described with a black-body spectrum with $kT < 0.4$\,keV and a
power-law spectral tail with photon index $\Gamma \sim 3.5$.  Such a
clear separation is consistent with predictions from the QED model by
\citet{Heyl07_london}.  However, the observed total and the pulsed
hard X-ray spectra are harder (index $\sim$1.0) than the photon index
of 1.5 predicted by this model.  Contrarily, the clear separation in
phase between the thermal and non-thermal pulses makes an
interpretation of the hard X-ray emission in terms of resonant
magnetic upscattering of the thermal soft photons by
ultra-relativistic electrons, accelerated along either open or closed
magnetic field lines unlikely \citep[proposed in the
  Compton-upscattering model by][]{Baring07_london}. At least part of
the thermal target photons should also arrive in the narrow phase
interval of the non-thermal pulse. This problem might be circumvented
if the thermal target photons are part of the isotropic thermal DC
emission, and the thermal pulse at phase 0.4 is not related to the
hard X-ray emission.

As mentioned before, \citet{Gotz07_1708} reported the detection of
correlated variability in hard X-ray flux and spectral index with
similar correlated variations at low-energy X-rays from \rxs1708. This
would imply correlated production processes at the same sites in the
magnetosphere. However, our analysis disproved the claimed variability
in the ISGRI-detected hard X-ray flux.

Very peculiar is the spectral shape of the component which contributes
significantly to the total-pulsed emission from \rxs1708 between 4 keV and
20 keV. It is not obvious what underlying emission processes can
produce such a spectral shape. This component exhibits the most narrow
pulse profile with a width of $\sim$0.25 in pulsar phase (FWHM, see
Fig.~\ref{fig:rxsmodpar}). It appears to be completely separated from
the soft pulse, but it is centered in phase on the hard X-ray pulse
which is about twice as broad. This means that the production site in
the magnetosphere of this curved component is confined to a narrow
region, possibly directed along open or the last closed magnetic-field
lines, similar to different dipole scenarios proposed for radio
pulsars \citep[e.g.][]{Ruderman75_polargap,Usov95_polarcap,
  Harding98_polarcaps,Arons83_slotgap,Muslimov03_slotgap,Cheng86_outergap,
  Romani96_outergap,Hirotani06_outergap}. Like is the case for radio
pulsars, this geometry appears to be very stable. The RXTE monitoring
observations proved that the pulse profiles and phase resolved spectra
were stable over 9 years. This might indicate that the non-thermal
emission is produced higher-up in the magnetosphere, above the region
closer to the neutron star with a strong  twisted (toroidal)
field.

The detailed results we obtained for AXPs \0142 and \rxs1708 clearly
prove that the total persistent high-energy emission from AXPs
consists of DC emission and pulsed emission exhibiting different
spectra \citep[already shown in][]{Kuiper06_axps}. Furthermore, the
pulsed emission consists of multiple components with different
spectra. As a consequence it does not seem meaningful to make detailed
model fits to just the total AXP spectra to derive a physical
interpretation of the emission. The physical origin and production
site of each different component has to be considered. Examples of
model fitting to the total X-ray spectra of magnetars are recent
applications of the resonant cyclotron scattering model first proposed
by \citet{TLK02}. Recent developments of this model can be found in
\citet{Lyutikov06_compton}, \citet{Guver06_stems,Guver07_1810} and
\citet{Fernandez07_cyclotron}. This model computes the result of
multiple scatterings of soft photons in the magnetospheres and was
developed to explain the apparently thermal black-body spectra with a
power-law extension to higher energies, as measured for energies below
$\sim$10 keV. In the extensive work by \citet{Rea08_cyclotron}
detailed fits with this model are merely made to the high-energy total
spectra of a comprehensive set of magnetars, including \rxs1708,
leading to interesting estimates of e.g. optical depths and
magnetospheric electron densities at the resonances. However, our
temporal and spectral decoupling of different components in the pulsed
emission (Figures \ref{fig:rxs10spec} \& \ref{fig:rxsmodpar}), shows
that between $\sim$4 keV and 20 keV a peculiar curved spectral
component contributes significantly to the pulsed emission, and is
unrelated to the soft thermal emission. This component contributes
also significantly to the total emission detected from \rxs1708 in the
energy interval where the resonant upscattering is effective. Hence,
it will obviously have an adverse impact on the parameter estimations
if this component has not been subtracted. On the other hand, the
resonant cyclotron-scattering model might offer a good explanation for
the soft pulse peaking at phase 0.4. Spectral modeling of the
corresponding spectrum, and separately of the thermal spectrum of the
DC emission will probe the model parameters in different sites in the
magnetosphere.

An important parameter to be reproduced by the models is the photon
index of the power-law spectral shape in the hard X-ray window and the
cut-off energy. The index $\Gamma \sim 1.0$ of the non-thermal pulse
and a spectral break at a few hundred keV is consistent with a
bremsstrahlung origin as first proposed in the corona model by
\citet{BT07}. But, further elaboration of the corona model by
\citet{Lyubarsky07_corona} lead to a different hard X-ray spectrum,
namely a superposition of bremsstrahlung radiation from the hot
atmosphere and Comptonization radiation from the extended corona.  The
latter produces a hard spectrum up to the MeV band and is inconsistent
with the COMPTEL upper limits at MeV energies. Furthermore, the
stability of the configuration over 9 years is a challenge for the
corona model. For the resonant upscattering model of
\citet{Baring07_london} the predicted photon indexes are considerably
softer and the spectra extend into the MeV range, possibly to
significantly higher gamma-ray energies. In this respect we should
emphasize that we donot see any indication for a break in the
\igr\ spectrum ($<$300~keV) of \rxs1708, more importantly, we donot
know for \rxs1708 and \0142 what the shapes of the spectra are above a
few hundred keV. Extreme cases are allowed by the data, e.g.  a hard
spectrum with index $\Gamma \sim 1.0$ with a sharp cut-off, as well as
with a break or bend with a flat spectrum extending underneath the
COMPTEL upper limits into the MeV or even GeV regions.

Similarly hard spectra with $\Gamma \sim 1.1$ have been measured in
the hard X-ray window up to 60--200 keV for several young
rotation-powered pulsars, e.g. AX~J1838.0-0655, PSR~J1846-0258,
PSR~J1811-1925, PSR~J1930+1852 and PSR~J1617-5055
\citep{Kuiper08_1838atel,Kuiper08_kes75}.  We also know that these
hard power-law spectra donot extrapolate to the MeV range (no
detections by COMPTEL).  Furthermore, those young pulsars detected
with COMPTEL in the MeV range, Crab \citep[for a multi-wavelength
  analysis similar to this work see][]{Kuiper01_crab} and PSR B1509-58
\citep{Kuiper99_1509}, have flat ($\Gamma \sim 2$) spectra at MeV
energies, and started with harder spectra at hard X-ray energies. In
addition, for three millisecond pulsars also an index $\Gamma \sim
1.1$ has been measured in the X-ray band up to ~25 keV \citep[PSR
  J0218+4232, PSR B1821-24 and PSR
  B1937+21;][]{Kuiper04_0218,Cusumano03_1937}. For one of these
millisecond pulsars, PSR J0218+4232, detection has been claimed even
between 300 MeV and 1 GeV with a soft spectral index of $\sim$2.6
\citep{Kuiper00_0218}, thus a drastic spectral bend or break is
required between the hard X-rays and the high-energy gamma rays.
These results on young as well as on old millisecond radio pulsars
prove that in dipole fields such hard non-thermal spectra can be
produced in stable configurations. The major difference with the AXPs
is that the energy source for ordinary pulsars is rotational
energy and for AXPs/magnetars magnetic energy.  It is interesting to
note that recently during a $\sim$50-day outburst magnetar-like bursts
have been reported from a very young rotation-powered pulsar (PSR
J1846-0258) with a strong magnetic field
\citep{Kumar08_kes75,Gavriil08_kes75bursts}.

A possible scenario for the persistent high-energy emission from AXPs
seems to emerge in which a soft thermal component, for \rxs1708
peaking at phase 0.4, originates in regions above and close to the
polar cap of the neutron star with resonant cyclotron scattering
offering a physical explanation. A large fraction of the thermal
photons will have a more isotropic distribution. These photons can be
responsible for the soft part of the DC emission spectrum, also with a
resonant cyclotron-scattering origin, but with e.g. different optical
depths and magnetospheric electron densities at the resonances. For
future studies, it will be interesting to accurately derive a spectrum
of just the DC emission for model fitting. Furthermore, the
non-thermal emission might be produced higher up in the outer
magnetosphere, where the magnetic field strengths is in the range
estimated for normal radio pulsars. The required strong current of
electrons and positrons can be sustained by the outward transfer of
magnetic helicity from the inner zone where it is injected.  Then the
acceleration of particles and the production of the observed
non-thermal photons has to occur in stable narrow zones along the
curved field lines to produce e.g. the narrow hard X-ray pulse seen
for \rxs1708. In the outer magnetosphere the scenario might resemble
those proposed for the high-energy emission from rotation-powered
pulsars. In fact, the approach taken by \citet{Baring07_london} is a
first attempt along these lines. However, they concluded that the
resonant magnetic Compton upscattering is most effective close to the
magnetar surface.

From only two of the 13 confirmed magnetars radio emission has been
detected and their radio emission properties have been measured in
great detail recently, notably from AXP \xte1810
\citet{Halpern05_1810,Camilo06_1810,Kramer07_1810} and AXP \e1547
\citep{Camilo07_1547,Camilo08_1547}. These very interesting AXPs show
transient radio emission, which is apparently related to the X-ray
variability \citep[e.g.][]{Ibrahim04_1810,Gotthelf05_1810,
  Gelfand07_1547,Halpern08_1547}.  These radio AXPs appear to be very
highly linearly polarized and have very flat spectra over a wide range
of radio frequencies, very similar characteristics to those of many
young ordinary radio pulsars. \citet{Camilo08_1547} note that, while
unexpected a priori, it appears that ideas developed to understand the
geometry of emission from ordinary pulsars along open dipolar field
lines also apply to known radio magnetars.  However, they also
conclude that both radio AXPs have also variable pulse profiles and
radio flux densities, not typical for young pulsars. 

Upon submission of this work, an extensive treatment appeared of pair
creation processes in an ultra-strong magnetic field and particle
heating in a dynamic magnetosphere \citep{Thompson08_radiomag1}, as
well as a self-consistent model of the inner accelerator with
implications for pulsed radio emission
\citep{Thompson08_radiomag2}. In the first paper Thompson considers
the details of QED processes that create electron-positron pairs in
magnetar fields. He discusses how the pair creation rate in the
open-field circuit and the outer magnetosphere can be strongly
enhanced by instabilities near the light cylinder and how this very
high rate of pair creation on the open magnetic field lines can help
to stabilize a static twist in the closed magnetosphere and to
regulate the loss of magnetic helicity by reconnection at the light
cylinder. In the second paper, one of the possible scenarios for
explaining the rapid radio variability, broad pulses and hard radio
spectra comprises a strongly variable current in the outer
magnetosphere and a high rate of pair creation sustained by a
turbulent cascade. It is interesting to note that stable geometries
are being considered to explain the production of transient radio
emission for AXPs involving high rates of pair creation on the open
magnetic-field lines in the outer magnetosphere. Such scenarios might
also be required to explain the production in the outer magnetosphere
on the orders of magnitude more luminous non-thermal X-ray emission.

\begin{acknowledgements}
We thank J\"orgen Kn\"odlseder for providing 
the SPI upper limits for \rxs1708.  This work is supported by NWO,
Netherlands Organisation for Scientific Research. The results are
based on observations with INTEGRAL, an ESA project with instruments
and science data centre funded by ESA member states (especially the PI
countries: Denmark, France, Germany, Italy, Switzerland, Spain), Czech
Republic and Poland, and with the participation of Russia and the
USA.  We are grateful to ASI, CEA, CNES, DLR, ESA,
INTA, NASA and OSTC for support. This research has made use of data
obtained through the High-Energy Astrophysics Center Online Service,
provided by the NASA/Goddard Space-Flight Center.
\end{acknowledgements}

\bibliography{../../4U0142/literature}
\end{document}